\documentclass[11pt, a4paper, twocolumn, copyright, gdm]{google}
\usepackage[authoryear, sort&compress, round]{natbib}
\usepackage{xcolor}
\usepackage{todonotes}
\usepackage{booktabs} 
\usepackage{tabularx} 
\usepackage{amssymb}  
\usepackage{float}
\usepackage{fancyvrb}
\usepackage{fvextra} 
\usepackage{booktabs} 
\usepackage{rotating} 
\usepackage{graphicx}
\usepackage{tcolorbox}
\usepackage{cleveref}
\tcbuselibrary{breakable}


\uselogo{} 

\title{Improving Group Deliberation through AI-Guided Self-Reflection}
\correspondingauthor{manonrevel@google.com}

\reportnumber{0001} 

\renewcommand{\today}{2026-10-26}

\author[1, 2, *]{Haiwen Li}
\author[1]{Crystal Qian}
\author[1]{Ben Usman}
\author[1]{Stephanie Chan}
\author[2, *]{Michiel Bakker}
\author[1]{Srivignesh Rajendran}
\author[1]{Avneesh Sud}
\author[1]{Thomas Leung}
\author[1]{Nikki Hariri}
\author[1]{John Canny}
\author[1, $\dagger$]{Nick Dufour}
\author[1, $\dagger$]{Manon Revel}

\affil[1]{\thepa{}{}}
\affil[2]{MIT}
\affil[*]{Work done at Google DeepMind}
\affil[$\dagger$]{Co-senior authors}

\begin{abstract}
As AI systems increasingly mediate human communication, identifying conditions under which they augment collective sense-making is a critical challenge. Through a randomized experiment ($N=1,173$ U.S. participants across $391$ small-group deliberations), we investigate whether AI-guided self-reflection improves subsequent deliberative discourse. We show that engaging with a non-directive AI chatbot instructed to structure self-reflection prior to group deliberation yields significantly higher deliberation quality (measured by the Discourse Quality Index, DQI) compared to a baseline of no self-reflection (Cohen's $d=0.31$) and a condition of unassisted self-reflection ($d=0.27$), and the positive impact is most pronounced in groups with disagreement. This improvement may be driven by more sophisticated argument justifications, greater consensus-building effort, and increased respect for counterarguments. Furthermore, after self-reflection, participants in the AI condition reported higher empathy toward dissenting peers and a deeper understanding of opposing reasoning. Beyond observed behavior, our findings are empirically relevant to open debates between political philosophy and psychology surrounding self-reflection, demonstrating that interactive AI could transform solitary thinking into productive internal reflection. Together, these results demonstrate that even simple AI-based reflective scaffolds can foster more productive group dynamics, opening new avenues for leveraging AI to augment collective intelligence. 
\end{abstract}

\begin{document}

\maketitle
\section{Introduction}

AI is increasingly shaping how people engage with contested issues. Recent work suggests that this influence can be constructive, through eliciting greater openness towards opposing views~\citep{lu2025ai}, reducing feelings of partisan animosity~\citep{thurgood2026human}, and durably changing conspiracy beliefs \citep{costello2024durably}. However, in these interventions, AI largely supplies the reasoning and the evidence that people then evaluate. Research on critical thinking suggests that such interaction mode may carry risks, linking reliance on AI to cognitive offloading and downskilling of human reasoning \citep{stadler2024cognitive,melumad2025experimental,gerlich2025ai}, though outcomes appear contingent on how systems are leveraged \citep{gerlich2025offloading,wang2025effect,lee2025impact,zhi2026investigating}. This raises the question of whether AI can support engagement by fostering human reasoning rather than replacing it.

Such support would be of particular interest in online deliberation, where discussions often fall short of mutual understanding and instead entrench polarization~\citep{fulay2026language}. To counter this dynamic, a host of approaches have been proposed and tested. Political theorists have highlighted self-reflection, or "deliberation within" \citep{goodin2000democratic,goodin2003does}, as a mechanism to encourage perspective-taking, surface gaps in one's own knowledge and beliefs, and foster empathy. Empirically, asking individuals to explain the mechanistic foundations of their positions has been found to expose an illusion of explanatory depth and moderates extreme views \citep{fernbach2013political}. However, solitary thinking has known limits: producing confirmatory arguments can reinforce one's existing beliefs, and reasoning works best when exercised in dialogue with others \citep{landemore2010talking, mercier2011humans}. A non-directive conversational AI that poses reflective questions could offer a middle ground between solitary reflection and AI that supplies reasoning and conclusions. 

Can AI-guided reflection help individuals unpack their reasoning, reproducing the benefits of deliberation within and, in turn, improve the quality of subsequent group deliberation? We explore deploying an AI Reflector (AIR) as a reflection scaffold prior to group interactions, akin to a \textit{reflective elicitation} in \cite{revel2026ai}'s framework. We examine whether AI can be designed to facilitate self-reflection while avoiding the pitfalls by focusing more on critically examining their own knowledge and reasoning and imagining the reasoning of others. Collaborating with Deliberate Lab \citep{qian2025deliberate}, we empirically test the impact of this pre-deliberation intervention on both individual- and group-level deliberative desiderata.

\subsection{Literature Review}
The idea of AI as reflectors \citep{revel2026ai} draws on research traditions at the intersection of philosophy, psychology and computer science.\footnote{Note that \cite{revel2026ai}'s AI Reflector is constituted of both reflective elicitation (our focus here) and synthesis generation (out of scope for this work).}

\textbf{Self-reflection.} A core tenet of this work is self-reflection -- the idea that self-generated insight can provide a constructive kind of knowledge. This concept takes roots in the Socratic method, where an interlocutor is guided toward self-generated insight, a practice Socrates conceptualized (and Plato recorded) as \textit{maieutik\={e} techn\={e}} or the ``midwifery craft'' \citep{jowett1894plato}. Building on these classical foundations, scholars have developed neo-Socratic frameworks. Most notably, the Nelsonian tradition \citep{nelson2025socratic,heckmann1981sokratische} departs from adversarial cross-examination in favor of non-directive facilitation, wherein participants reflect collaboratively upon shared personal experiences. In Nelsonian dialogue, a facilitator assists participants in identifying the boundaries of their own comprehension through concrete, real-world examples~\citep{saran2004enquiring}. This non-directive ethos was subsequently operationalized and evaluated within clinical and applied psychology through Motivational Interviewing \citep{miller2012motivational,lundahl2010meta,rubak2005motivational}, as well as within philosophical frameworks of critical thinking \citep{virtanen2026critical}. Further afield,  Goodin has explored how self-reflection can play a vital role in democratic deliberation practices\citep{goodin2000democratic}. His ``deliberation within'' framework posits that the internal cognitive process of processing new information,  weighing competing considerations and perspective-taking could serve as a foundational component of improved democratic discussions.

Despite these benefits, solitary reflection is also susceptible to systematic biases: according to the argumentative theory of reasoning \citep{mercier2011humans,mercier2012reasoning}, human reasoning evolved largely to produce and evaluate justifications within social argumentation. Because this mechanism is fundamentally designed to defend one's stance against others, when individuals reason in isolation, they lack the natural corrective of social opposition, predisposing them to recruit evidence that selectively defends their prior convictions. Consequently, solitary contemplation could risk entrenching existing attitudes rather than refining them, underscoring that unstructured self-reflection is not an unalloyed mechanism for depolarization. 

This risk is also compounded by the ``Illusion of Explanatory Depth'' (IOED), wherein individuals systematically overestimate their knowledge in an issue \citep{rozenblit2002misunderstood}. When reflection is left unstructured, it typically takes the form of rehearsing reasons for one's position and reinforcing one's stance. However, \citet{fernbach2013political} demonstrated that prompting individuals to articulate explicit, step-by-step causal mechanisms (as opposed to enumerating justifications) reduces IOED, thereby lowering attitude extremity and increasing intellectual humility regarding complex policy issues. Consistently, empirical investigations of citizen juries demonstrated that participants' attitudes undergo more substantial shifts during individual information-gathering and reflective stages than during formal collective discussions \citep{goodin2003does}. Moreover, reflective dispositions such as perspective-taking~\citep{galinsky2005perspective,todd2011perspective} and cognitive humility~\citep{leary2017cognitive,krumrei2016development} have been found to foster more respectful groups~\citep{liht2013preventing,porter2018intellectual,peracha2022promoting}.

\textbf{AI and deliberation.} 
As an idea, AI reflectors are situated within the growing corpus examining AI in deliberative settings. Prior paradigms, such as the Habermas Machine \citep{bakker2022fine,tessler2024ai}, have focused on real-time mediation across group participants to forge consensus. Personalized one-on-one dialogues have been shown to durably reduce conspiracy beliefs \citep{costello2024durably,boissin2025dialogues,costello2026large}. Socratic-style interaction has been found to support individual judgment and sharpen logical discernment\citep{danry2023don, park2023thinking}. Relatedly, reflection nudges have been shown to improve the deliberativeness of individual online contributions \citep{yeo2024help}, but not of real discussion among people. In general, these benefits have been assessed predominantly through self-reported, individual-level measures, leaving open whether AI-guided reflection translates into concrete downstream behavior. Our approach diverges from these paradigms by using the AI as a non-directive, pre-deliberative reflective scaffold to prepare humans for real-time, human-to-human discussion. Crucially, our setup is not replacing human deliberation with AI; rather, it explores whether certain kinds of AI interaction can improve later human dynamics.

\subsection{Paper Overview}
To evaluate the impact of pre-deliberation AI reflection on subsequent group deliberation, we designed a randomized controlled experiment comprising three experimental conditions: (1) a baseline control condition with no structured reflection, (2) an unassisted, static reflection condition, and (3) a self-reflection condition with AI reflector (AIR). 

Across all conditions, participants first review a standardized policy primer covering one of three real-world policies and rate their initial stance. In the control condition, participants are guided to watch a 7-minute YouTube video. In the static, human-only condition, participants follow text-based reflection instructions. In the AIR condition, participants engage in an interactive dialogue with AI to reflect. Finally, participants join a synchronous, anonymous group deliberation to discuss the policy outlined in the initial primer.

Our primary outcome of interest is deliberative quality, operationalized using an automated implementation of the Discourse Quality Index (DQI; \cite{steenbergen2003dqi}) evaluated via a large language model. The DQI assesses seven deliberative dimensions: general conversational engagement, level of justification, content of justification, respect toward groups, respect toward demands, respect toward counterarguments, and constructive consensus-building efforts. In addition to textual deliberation dynamics, we collect longitudinal, self-reported psychometric measures across stages of the experiment to capture shifts in stance certainty, preparedness, mechanistic understanding, and empathy.

The remainder of this paper is structured as follows. In Section~\ref{sec:methods}, we introduce the AI reflector, the experiment, the DQI measure, and analysis methods. Section~\ref{sec:results} presents the primary and exploratory empirical findings, analyzing both the effects on group deliberation quality and participants' self-reported measures. Finally, Section~\ref{sec:discussion} discusses the theoretical and practical implications of our findings for human-AI interaction and online democratic spaces, addresses limitations, and outlines future avenues for non-directive algorithmic facilitation.
\section{Methods}\label{sec:methods}
\begin{figure}
    \centering
    \includegraphics[width=\linewidth]{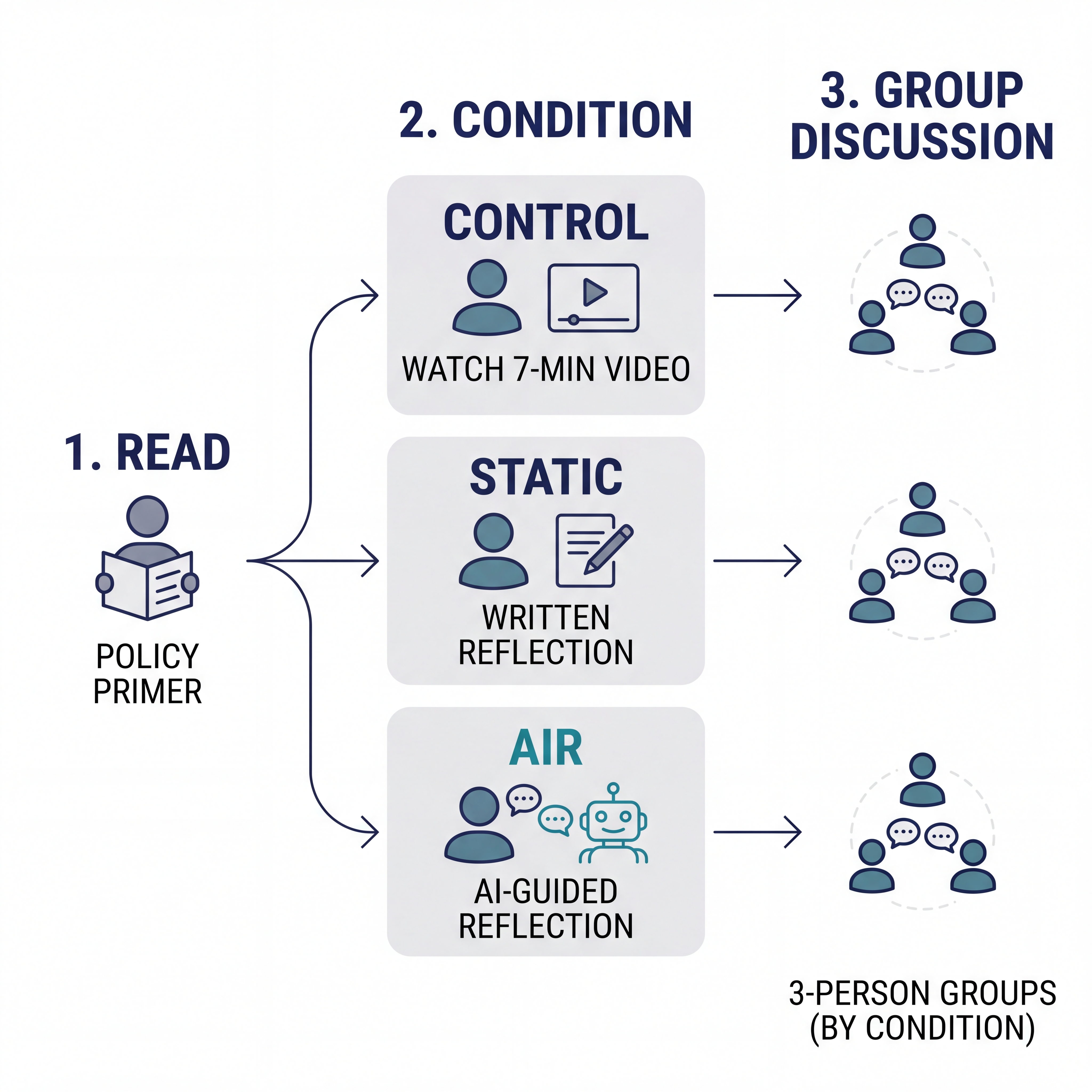}
    \caption{\textbf{Experimental Design.} Participants were assigned to a single policy and experimental condition. The experimental flow involved reading a policy primer, completing a reflection exercise, and joining a three-person group discussion about the policy. Image generated with Nano Banana.}
    \label{fig:experiment_design}
\end{figure}

\subsection{Experimental Conditions}

\paragraph{AI Reflector (AIR):} Participants engage in a 7-minute conversation with an LLM explicitly prompt-engineered to serve as a non-directive, neo-Socratic reflective scaffold \citep{revel2026ai}. We use an open-source model, \texttt{gemma-4-31b-it}, and set the reasoning level to ``High'' for the AI reflector. \footnote{The model does not have web search access, which is unavailable in Deliberate Lab. Instead, when participants explicitly ask for factual information, AIR is instructed to prompt the participants to search by themselves.}

AIR's system prompt directs it to act as a strictly neutral reflection partner\footnote{See \Cref{app:system_prompt} for more details about the prompt and \Cref{fig:experience-air} for self-reported experience with the AIR system.}, helping users articulate their core views, distinguish their factual knowledge from underlying assumptions, and constructively explore opposing perspectives without feeling attacked or pressured to change their minds. To ensure a comfortable and natural dialogue, the prompt imposes a 2–3 sentence response limit, discourages conversational filler or repetitive questioning, and incorporates time-bound pacing. See prompt used in \Cref{app:system_prompt}. 

\paragraph{Static:} Participants complete a 7-minute unassisted, individual reflection exercise under static instructions, writing a short statement that articulates their stance without algorithmic interaction.

The Static reflection instruction is designed as a self-guided writing prompt that directly mirrors the core cognitive exercises of the AIR condition. It translates AIR's prompt into three open-ended reflection questions: it asks the participant to articulate their current view; evaluate what they factually know, what they do not know, and what they are assuming; and prompts the participant to consider how someone with an opposing view might reason. See instruction used in \Cref{app:static_prompt}

\paragraph{Control (No Self-reflection):} As a baseline, we use a no-reflection condition where participants watch 7-minute YouTube video\footnote{https://www.youtube.com/watch?v=jCKt02NGjfM}, that is irrelevant to the policy of interest, before joining the group deliberation. This irrelevant video task serves as a placebo to standardize the pre-deliberation time delay.

\subsection{Experimental Design and Procedure}
We conducted the experiment, sketched in Figure ~\ref{fig:experiment_design}), using Deliberate Lab \citep{qian2025deliberate}. Participants entered the study and were randomly assigned to one of the experimental conditions. First, they reviewed a brief, neutral policy primer to learn about basic facts relevant to the assigned policy, and then reported their initial policy stance ($T_1$).\footnote{To prevent the survey questions from acting as reflection nudges prior to the intervention, this pre-reflection survey only asked participants to report their initial policy stance.} Next, they transitioned to the pre-deliberation intervention where they engaged in their assigned reflection activity—either watching an irrelevant video (Control), completing a static unassisted reflection, or interacting with the AI Reflector. Immediately following this, they completed a post-reflection survey ($T_2$) measuring immediate cognitive, attitudinal, and affective shifts before entering a waiting room to be grouped. Finally, participants were matched into three-person groups from the same condition for a synchronous group deliberation, engaging in a text-based discussion on the policy. The study concluded with a post-deliberation survey ($T_3$) to capture downstream attitudinal changes, group discussion dynamics, and feedback on the AI Reflector.

\paragraph{Policy Topics} We selected three contemporary, controversial United States public policy topics, adopted from \citet{akbulut2026evaluating}: (1) Ending Federal Public Broadcasting Funding, (2) Medicaid Work Mandates, and (3) Using Live Facial Recognition Technology in Public by Law Enforcement. Each participant is assigned to a single policy topic. A complete description of the policies and the primers is available in \Cref{app:policy_primers}.

\paragraph{Participant Recruitment and Sample Size} The experiment is approved by Google DeepMind Human Behavioural Research Ethics Committee (HuBREC 26 002). We recruited $N = 1,465$ participants via Prolific, of whom $1,253$ reached the group-formation stage, and $1,173$ were successfully matched into active discussion groups where everyone completed the study. This final sample has a median age of 34 and $59.6\%$ female. There's no differential attribution across conditions ($\chi^2(2)=1.111,p=0.574$). This sample size yielded $391$ groups total, with $128$ groups in the AIR condition, $132$ in the Static reflection condition, and $131$ in the Control condition. An \emph{a priori} power analysis indicated that this sample size provides over 80\% power to detect small-to-moderate main effects ($d \approx 0.30,$ where $d$ is the Cohen's d effect, the mean difference between two groups per pooled standard deviation) at $\alpha = 0.05$. Participants were compensated at an hourly rate of \$20.00 with a \$2.00 bonus\footnote{Engagement bonus was awarded to participants whose group deliberation message volume were among the top 25\% in each condition. Participants were not informed of how the engagement is measured.} for an average session time of 30 minutes.

To avoid ideological homogeneity, we utilized participants' $T_1$ stances to construct three-person groups containing at least one proponent and one opponent of the policy. For the federal broadcasting policy and the public facial recognition technology policy, due to heavy skew in the recruitment sample, this matching constraint was relaxed to permit one-sided groups. In our final sample, $16.80\%$ of groups discussing live facial recognition technology and $52.80\%$ of groups discussing federal funding for public broadcasting failed to include participants with opposing views. We present our analyses aggregated across all groups, across the two-sided groups, and across the one-sided groups.

\subsection{Outcome Measures}
\paragraph{Discourse Quality Index on Deliberation Transcripts}To evaluate our primary outcome of group deliberation quality, we adapt the Discourse Quality Index (DQI) from \cite{steenbergen2003dqi}. Grounded in Habermasian discourse ethics, the DQI is a widely used coding scheme for measuring quality of political deliberation. We lightly modify the coding scheme to fit synchronous online chat by replacing the parliamentary ``participation'' dimension with conversational ``engagement'' \citep{stromer2007measuring}. The resulting index scores transcripts across seven dimensions: level of justification, content of justification, respect toward groups, respect toward demands, respect toward counterarguments, constructive consensus-building efforts, and general conversational engagement.

Transcripts are scored at the message level using an LLM-as-a-judge framework with \texttt{gemma-4-31b-it} configured for high reasoning effort. The LLM judge first evaluates whether a message is relevant to the policy discussion. Following \cite{steenbergen2003dqi}, if a message contains a relevant part, it is coded for its discourse quality. Each message is scored across three independent seeded runs. We establish relevance via majority vote across seeds. For relevant messages, sub-dimension scores are averaged (round-half-up) across runs and summed to produce a composite message score. Our primary group-level DQI metric is the average of these composite scores across messages within a discussion group. Inter-rater agreement (benchmarked against \texttt{gemini-3.1-pro-preview}), and manual human-validation checks (conducted on a random sample of 108 messages) confirmed high measurement consistency and reliability (see Appendix \ref{app:dqi_robust} for full agreement statistics). Prompts and rubrics are detailed in Appendix \ref{app:dqi_prompt}. 

\paragraph{Self-reported Survey Measures}We capture longitudinal cognitive, attitudinal, and affective shifts through survey batteries administered at $T_1$, $T_2$, and $T_3$. These batteries measure: self-reported policy stance ($T_1, T_2, T_3$), stance certainty ($T_2, T_3$), practical policy understanding ($T_2, T_3$), perceived conversational preparedness ($T_2, T_3$), empathy and perspective understanding toward opponents ($T_2, T_3$), post-deliberation evaluations of experience of group deliberation and AI reflection ($T_3$). The complete text and scale details for all survey items are provided in Appendix \ref{app:survey_items}.

\subsection{Statistical Analysis}

To estimate the causal effect of AIR on our primary outcome of subsequent group deliberation quality, we regress group-level DQI on treatment assignment, group initial stance composition, and aggregated demographics (age, gender, ethnicity, employment status, student status):
\begin{align*}
    \mathit{DQI}_{gp} &= \alpha_p + \beta * \mathit{Treatment}_g + \gamma_1 * \mathit{PropSupport}_{gp} +\\
    &\gamma_2 * \mathit{GroupDemographics}_g + \epsilon_{gp}
\end{align*}
where $DQI_{gp}$ is the average discourse quality for group $g$ on policy $p$, $Treatment_g$ is the assigned condition, $\alpha_p$ denotes policy fixed effects, $PropSupport_{gp}$ is the centered proportion of group members supporting the policy at baseline, and $GroupDemographics_g$ is a vector of group-level demographic shares. We estimate heteroskedasticity-robust standard errors (HC3) to account for non-constant error variance from highly volatile group-level dynamics.

Individual-level survey items are analyzed using separate regressions controlling for treatment indicators, initial stance, policy fixed effects, demographic covariates, and group random intercepts (when applicable).

Our confirmatory analysis focuses on the three pairwise comparisons of group-level DQI across our experimental conditions: (1) AIR vs. Static, (2) AIR vs. Control, and (3) Static vs. Control. These comparisons are estimated as linear contrasts of the treatment coefficients and tested with Wald tests using the model's coefficient covariance. We control the False Discovery Rate (FDR) over the primary family of three pairwise group-level hypotheses by applying the Benjamini--Hochberg (BH) procedure. We report both unadjusted (raw) p-values and BH-adjusted p-values for these comparisons. Exploratory and secondary analyses are presented with unadjusted p-values. Additionally, we report all effects in terms of Cohen's $d$ throughout the manuscript for ease of interpretation.

\section{Results}\label{sec:results}
This section outlines the impact of AIR on group deliberation quality, operationalized by DQI, and on self-reported attitudes. We find that AIR improves DQI over both the Static and Control conditions, and the effect is driven by two-sided groups. Further, AIR has a positive impact across all participants' DQI, while the Static condition selectively improves deliberation quality in the messages of those most prepared. Next, participants who engaged with AIR self-reported higher empathy and understanding for the other side. Last, we find that the timing of opinion change happens after self-reflecting in AIR while it takes place after deliberating in the other conditions. 
\subsection{AIR Improves Deliberation Quality}
\begin{figure*}[h!]
  \centering
  \includegraphics[width=\textwidth]{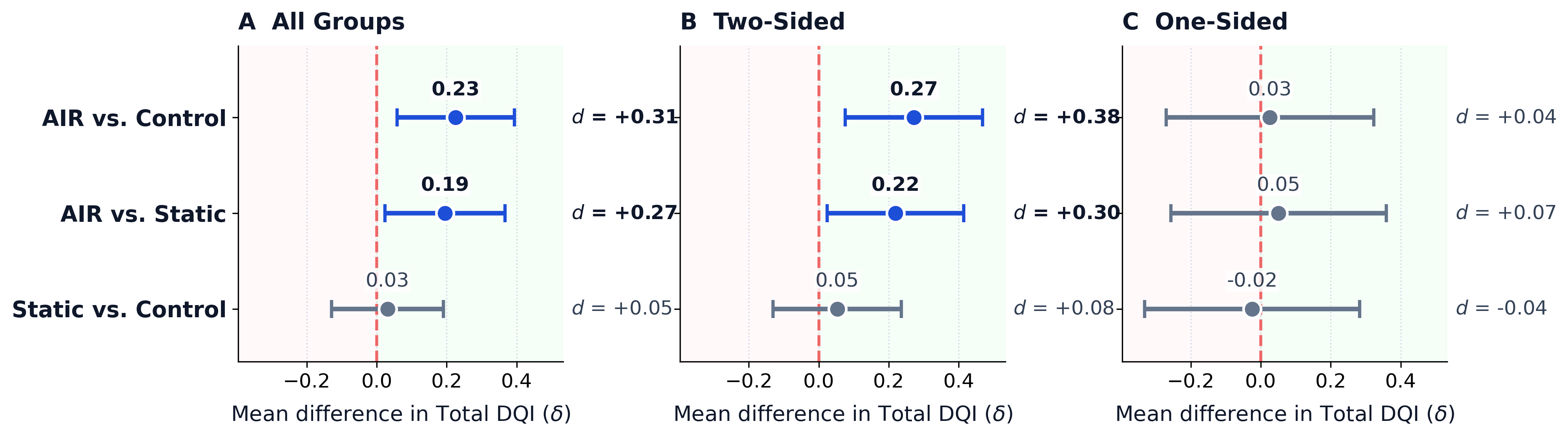}
\caption{\textbf{Treatment effects on group-level Total DQI overall and by group polarization.}
  Panels show pairwise treatment contrasts across (\textbf{A}) all groups, (\textbf{B}) two-sided groups, and (\textbf{C}) one-sided groups. Points and horizontal bars show model-adjusted mean differences ($\delta$, in DQI points, printed above each marker) with 95\% confidence intervals; standardized effect sizes (Cohen's $d$) are reported to the right of each panel. The dashed vertical line marks $\delta = 0$. Blue points and bold labels indicate significance at $\alpha = .05$. We use BH-adjusted $p$-values for the three contrasts with \textit{all groups} (primary). }
  \label{fig:dqi_main2}
\end{figure*}

\paragraph{Group-level DQI} Figure~\ref{fig:dqi_main2} presents the comparison of deliberative quality across the three conditions using the total, group-level DQI (individual-level plot can be found in \Cref{app:dqi-participants}). After self-reflection with AIR, group deliberation quality was scored significantly higher than the groups in the Control condition, with a small-to-moderate size effect ($d = 0.31$, $p = .009$, $p_{\text{adj}} = .026$). To build intuition, this means that $61.8\%$ of the groups in the AIR arm score above the median group in the Control condition. In contrast, Static condition yielded no detectable improvements over the Control condition. The direct comparison between the two reflection formats also favored AIR over Static reflection. AIR groups outperformed Static groups by a small-to-moderate margin ($d = 0.27$, $p = .026$, $p_{\text{adj}} = .039$).

To isolate the specific drivers of this improvement, we conducted an exploratory analysis across the seven sub-dimensions of the DQI (see \Cref{app:dqi-subdimension} for detailed results). Compared to the Control condition, AIR groups showed positive trends in several areas: respect for counterarguments ($d = 0.20$, $p = .101$), consensus-building efforts (group: $d = 0.21$, $p = .078$), and level of justification (group: $d = 0.22$, $p = .077$). When compared directly to Static reflection, AIR prompted participants to provide significantly more sophisticated justifications ($d = 0.27$, $p = .035$). Static reflection did not significantly alter any sub-dimension relative to the Control (all $p > .25$). 

In the discussion, these sub-dimensions distinguish brief, positional statements from substantive deliberation. Messages with low scores on these dimensions and the DQI merely state bare positions, such as ``\textit{I strongly agree with the policy}'' or ``\textit{I'm against it}''. By contrast, high-scoring turns in AIR typically integrate multiple dimensions within a single contribution. For example, when talking about facial recognition technology, one participant said ``\textit{These advantages do exist, but live public scanning is too wide a dragnet. Tools that are targeted, such as post-incident forensics searches backed by warrants, do the same job without making public places constantly surveilled.}'' Similarly, in the Medicaid discussion, a participant bridged work requirements and health limitations by proposing tiered hours for vulnerable recipients: ``\textit{I do think that allowing school or community service can also help. That can be less taxing depending on what it is. I think [the current work hour requirement] is kind of hit or miss. Maybe 10--15 for people who are sick or injured of low impact volunteering}''. To further illustrate how these mechanisms unfold in practice, Appendix~\ref{app:qualitative_examples} presents qualitative trajectories comparing how solitary static writing versus interactive AIR dialogue steered participants from their initial thoughts to their subsequent justifications in the group discussion. 

\paragraph{Heterogeneity across Group Composition} Recall that, due to highly skewed opinion distributions in the recruited sample, several groups lacked ideological diversity. To evaluate how this affected our findings, we examined the heterogeneity of the treatment effects by distinguishing \textit{two-sided} groups (whose members held opposing pre-reflection stances; $n = 303$ groups, $909$ participants) from \textit{one-sided} groups ($n = 88$ groups, $264$ participants). We interacted the treatment indicators with a dummy variable for one-sided groups and calculated subgroup-specific contrasts (see the middle and right plot in Figure~\ref{fig:dqi_main2}).

We observe the benefits of AIR heavily concentrated within the two-sided groups. Within the groups where people disagreed with each other, AIR substantially raised deliberative quality compared to the Control group level ($d = 0.38$, $p = .007$) with larger effect sizes. AIR also significantly outperformed Static reflection ($d = 0.30$, $p = .028$), while Static reflection still did not differ from the Control group ($d = 0.08$, $p = .569$). Sub-dimension analyses within the two-sided groups mirrored these main findings, with the level of justification, respect for groups, and consensus-building efforts driving the gains.

Conversely, in groups lacking opposing pre-reflection stances, deliberative quality remained similar across all three conditions. These one-sided groups also exhibited lower deliberation quality overall.  

\paragraph{Differential Impacts on DQI} An examination of the individual-level DQI distribution across participants shows that Static reflection primarily shifts the upper quantiles ($Q_{75}$--$Q_{90}$) while leaving the lower half and median close to Control, whereas AIR shifts the speaker distribution upward across all quantiles (\Cref{fig:individual-dqi-percentile-shift} in \Cref{app:dqi-participants}). Interestingly, this hints at AIR potentially lifting up all participants' deliberation preparedness, while Static reflection seems to only help those with the highest baseline. 
\subsection{AIR's Impact on Self-Reported Attitude}
\begin{figure*}[h!]
  \centering
  \includegraphics[width=\textwidth]{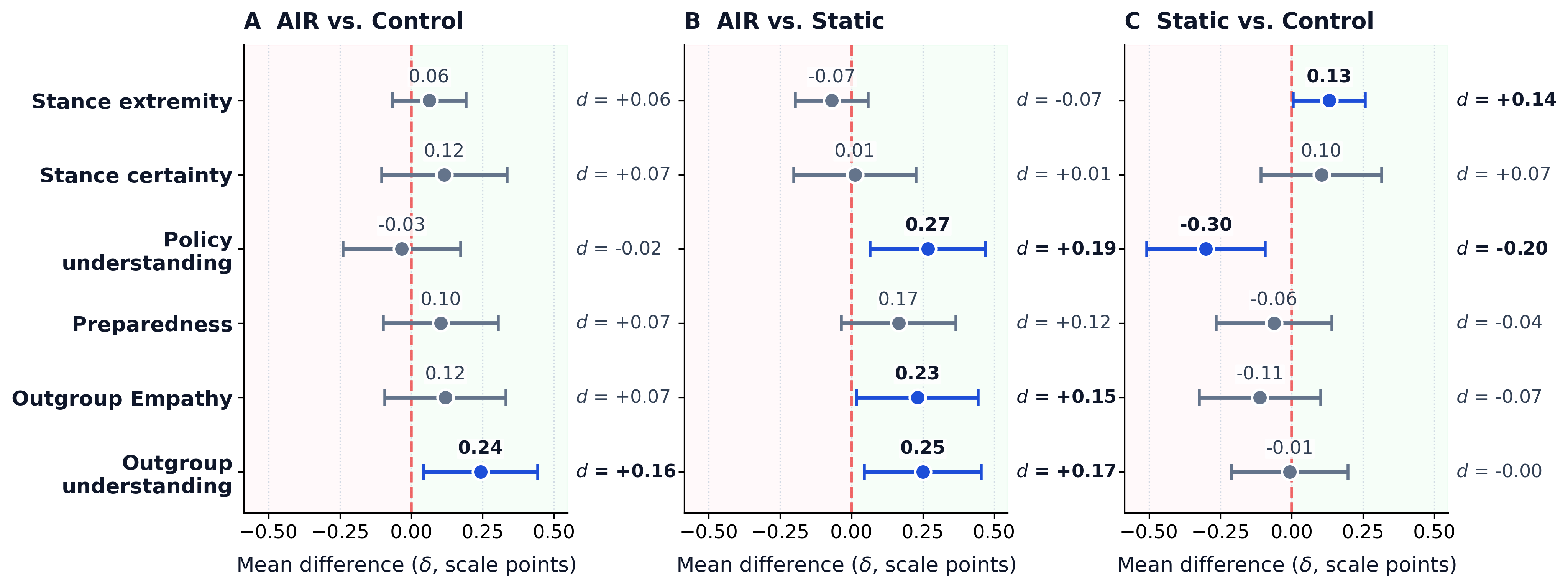}
\caption{\textbf{Treatment effects on post-reflection, pre-discussion (T2) survey outcomes.}
  Panels show pairwise treatment contrasts (\textbf{A}: AIR vs.\ Control; \textbf{B}: AIR vs.\ Static; \textbf{C}: Static vs.\ Control) across the six T2 survey outcomes (1--7 Likert scale; 0--3 folded scale for stance extremity). Points and horizontal bars show model-adjusted mean differences ($\delta$, in scale points, printed above each marker) with 95\% confidence intervals; standardized effect sizes (Cohen's $d$) are reported to the right of each panel. Blue points and bold labels indicate significance at $\alpha = .05$ using unadjusted $p$-values (exploratory).}
  \label{fig:survey_t2}
\end{figure*}

Figure~\ref{fig:survey_t2} illustrates participants' self-reported attitudes captured after the reflection/video-watching stage but prior to group deliberation, revealing the immediate impact of reflection. Self-reflection with AIR primarily changed how participants see the opposing side. Compared to Static reflection, participants who engaged with AIR reported significantly higher empathy toward those holding opposing views ($d = 0.15$, $p = .033$) and a better understanding of their underlying reasons ($d = 0.17$, $p = .017$) with small effect sizes.\footnote{While small, these effects could have significant compounding effects at the scale of, for instance, a large online forum.} AIR also improved participants' understanding of the opposing side's reasoning relative to the Control condition ($d = 0.16$, $p = .017$). Static reflection did not statistically differ from the Control condition on either of these out-group measures. 

Conversely, Static reflection lowered participants' self-reported understanding of the policy compared to the two other conditions (vs. Control: $d = -0.20$, $p = .005$; vs. AIR: $d = -0.19$, $p = .010$). This self-reported decline aligns closely with the reduction of the IOED, indicating that our implementation of Static reflection successfully replicated established cognitive-psychological phenomena observed in previous literature \citep{fernbach2013political}. However, this solitary exercise also came with a trade-off: Static reflection produced more extreme policy stances than the Control condition ($d = 0.14$, $p = .041$). This polarization trend supports existing theoretical assertions that solitary, monological thinking may reinforce one's prior beliefs \citep{landemore2010talking,mercier2012reasoning}.

After the group deliberation (T3) AIR's increase in outgroup empathy and understanding, while still having positive magnitude, are no longer significantly greater than the other two conditions. Three significant differences remained at T3: Static reflection condition participants rated their understanding of the policy lower
than Control ($d = -0.21$, $p = .005$), AIR participants felt more prepared than Static participants ($d = 0.15$, $p = .039$) to discuss about the policy with others, and AIR participants reported higher outgroup empathy than Static participants ($d=0.15$, $p=0.036$). But overall, participants in all three conditions ended up at similar levels on most outcomes after group deliberation. We speculate that AIR reflection provokes internal simulation of deliberation (\cite{goodin2000democratic}'s ``deliberation within'', leading participants to gain part of what group deliberation offers before the discussion began (which may have lead to the increased in deliberation quality in this condition). Once participants in the other two conditions took part in an actual group discussion and had a comparable deliberative experience, the conditions converged.

\subsection{AIR's Impact on Opinion Changes}
\paragraph{Opinion change trajectory}Evaluating the overall directional trajectory of participants' policy attitudes across the entire experimental pipeline ($T_1 \to T_3$), we see that participants across three conditions on average moved toward opposing both Facial Recognition Technology ($\Delta = -0.34, p < .0001$) and Medicaid work requirements ($\Delta = -0.33, p < .0001$), while attitudes toward ending federal funding for public broadcasting remained stable ($\Delta = +0.05, p = .53$). The direction of stance change doesn't differ across conditions.

Zooming in, when examining the total magnitude of individual opinion shift from baseline to post-deliberation ($T_1 \to T_3$), overall movement is virtually identical across all three experimental arms (AIR: $\Delta = 0.73$; Static: $\Delta = 0.76$; Control: $\Delta = 0.79$; all pairwise $p \ge .43$). However, dis-aggregating by stage reveals that the time at which these shifts occurred differed by condition. During the reflection phase ($T_1 \to T_2$), self-reflection with AIR induced more individual reconsideration than both the Static and Control conditions. Specifically, $30.1\%$ of AIR participants shifted their 7-point Likert agreement score, and $13.8\%$ switched their categorical stance (Pro, Neutral, or Against), compared to $23.0\%$ and $7.4\%$ in the Static arm and $17.4\%$ and $6.4\%$ in the Control arm. Similarly, the mean absolute stance shift on the 7-point scale between $T_1$ and $T_2$ was significantly larger in the AIR condition ($\Delta = 0.42$) than in both Static ($\Delta = 0.30$; AIR - Static: $d=+0.25, p=0.010$) and Control ($\Delta = 0.24$; AIR - Control: $d+0.18, p<0.001$). Conversely, Control participants exhibited larger shifts during the group discussion itself ($T_2 \to T_3$: $\Delta = 0.70$ vs.\ $\Delta = 0.55$ in AIR; AIR - Control: $d=-0.15, p=0.044$), implying that AIR front-loads individual attitude updating prior to group interaction (see Figure~\ref{fig:opinion_change_3intervals} for more details). Crucially, we do not treat belief change as normatively desirable or undesirable in itself. Reflection-induced opinion shift could reflect either constructive epistemic reflection as participants weigh competing perspectives and refine their initial view, or undesirable AI steering. However, the fact that the overall opinion change ($T_1 \to T_3$) does not differ between AIR and the other two conditions suggests that AIR does not systematically steer or over-influence participants.

\paragraph{Opinion convergence} Another noteworthy result is the impact of deliberation on opinion convergence, which we operationalized as the change in the distance between the most extreme opinions within each group from $T_2$ to $T_3$. We find that while actual discussion narrows opinion variance across all conditions, the magnitude of opinion convergence is strongest among those who prepared via AIR. Groups in the Control condition built moderate consensus ($\Delta = -0.37, p = .008$), while groups in the Static ($\Delta = -0.45, p < .001$) and AIR ($\Delta = -0.56, p < .001$) conditions achieved progressively tighter opinion convergence. This suggests that AIR reflection may prepare participants to build deeper consensus when they subsequently deliberate with others.
\section{Conclusion}\label{sec:discussion}
In this study we investigated whether AI-guided self-reflection can serve as an effective pre-deliberative intervention to enhance subsequent group policy discourse. Our findings speak to long-standing debates regarding the efficacy of internal contemplation in democratic life. Although static reflection successfully reduced participants' overconfidence in policy knowledge, such epistemic humility failed to translate into improved deliberation. In contrast, by engaging participants in interactive Socratic dialogue, AIR supplied the dynamic epistemic friction that solitary reflection lacks, boosted out-group empathy and understanding, and fostered higher quality deliberation. Overall, these results provide empirical grounding for the normative framework around AI Reflectors proposed by \cite{revel2026ai} and highlight how the Illusion of Explanatory Depth, the argumentative theory of reasoning, and ``deliberation within'' can fit into a cohesive, complementary theory.

A limitation of our design is that all participants within a given group were from the same experimental condition. This is unrealistic in real-world civic environments, where universal participation in reflection is rare due to varying levels of civic interest, time constraints, and compliance. Extending the current work to heterogeneously treated groups would clarify whether the benefits of individual reflection can generate positive peer spillovers. If an individual enters a discussion with well-substantiated justifications and engage constructively with counterarguments, they may elevate the epistemic standards of the entire group. From an implementation perspective, establishing that reflection produces positive externalities across social networks would suggest that population-level improvements in collective reasoning could be achievable even with engaging only a fraction of the people. 

Our findings also bear directly on the design of conversational human-AI interaction. AIR models an interaction paradigm in which the AI acts as an active sounding board that occasionally asks probing questions, providing a cognitive scaffold for self-reflection. Importantly, we emphasize that our results provide a proof-of-concept for this paradigm and do not represent off-the-shelf LLMs: we designed AIR to be neutral and non-directive, while standard commercial LLMs are purposed to answer questions.

This reflective paradigm could carry practical implications for online public spheres. Embedding lightweight, conversational reflection prompts directly into online platforms—such as Reddit or X—could introduce productive cognitive friction into everyday digital discourse. Even a brief, 30-second micro-intervention prior to submitting a comment may interrupt reflexive, emotionally charged posting behaviors and encourage users to articulate their reasoning and consider counter-perspectives before broadcasting their views.

\bibliography{main}
\onecolumn
\appendix
\section{Generative AI Usage Statement}
Various Gemini and Google models were used to support multiple stages of the research. We used these models to identify gaps in our literature review, assist with data analysis and visualization, and revise manuscript from human-provided feedback. The authors conducted verification of all generative AI outputs and extensively edited all drafted text. The authors remain fully responsible for the correctness, originality, and integrity of all content in this work.

\section{Ethics and Privacy Statement}

Our study demonstrates that the AI Reflector (AIR) significantly improves group deliberation quality without causing disproportionate overall opinion-changing compared to human-only reflection or no reflection. However, the risk still remains that even a carefully engineered "neutral" AI could subtly and unintentionally steer public opinion. While we did not observe such biasing effects with the policy tested in our study, this does not rule out any future risk, and we emphasize the need for open-source prompt transparency and further auditing of AI-guided interventions.

This study was approved by Google DeepMind Human Behavioural Research Ethics Committee (HuBREC 26 002), with all participants recruited via Prolific providing informed digital consent. All Prolific participant data and conversation transcripts were collected anonymously and were reported solely in anonymized formats.

\section{Detailed Experimental Protocols and Materials}
\subsection{Detailed Phase Breakdown}
The experimental protocol unfolds sequentially across six distinct phases:
\begin{itemize}
    \item \textbf{Phase 1: Policy Primer.} Participants review a standardized, neutral background document detailing the policy topic under consideration to establish common factual grounding.
    \item \textbf{Phase 2: Pre-Reflection Survey.} Participants report their initial stance on the policy.
    \item \textbf{Phase 3: Pre-Deliberation Intervention.} Participants engage with their assigned condition: watching an irrelevant video without reflection activity (Control), completing an unassisted reflection following static instructions (Static Reflection), or completing an interactive dialogue with the neo-Socratic reflective LLM (AIR).
    \item \textbf{Phase 4: Post-Reflection Survey.} Participants respond to survey questions evaluating their stance, stance certainty, policy understanding, preparedness, outgroup empathy, and outgroup understanding. After this survey, participants enter a waiting room to match with a group.
    \item \textbf{Phase 5: Synchronous Group Deliberation.} Three-person groups from the same experimental condition interact in an unmediated, synchronous, text-based discussion room to deliberate on the policy topic.
    \item \textbf{Phase 6: Post-Deliberation Survey.} Participants complete a survey with the same questions as the post-reflection survey to capture downstream shifts, and answer additional questions on the group deliberation experience, alongside evaluations of the AI reflector.
\end{itemize}
\label{app:survey_flow}
\begin{table}[H]
\centering
\small
\caption{Pre-Deliberation and Discussion Instructions Administered to Participants}
\label{tab:participant_instructions}
\begin{tabularx}{\textwidth}{llX}
\toprule
\textbf{Experimental Condition} & \textbf{Stage} & \textbf{Instructions and Prompts Displayed to Participants} \\
\midrule
\textbf{AI Reflection} & Interactive Reflection & For the next \textbf{7 minutes}, you will reflect on your opinion with an AI chatbot regarding the following policy: \textit{\{\{policy.text\}\}}. As you reflect, consider:
\begin{itemize}
    \setlength{\itemsep}{0pt}
    \item What is your current view, and what is the main consideration behind it?
    \item What do you factually know about the policy, what do you not know, and what you may be assuming?
    \item How might someone with a different view argue about the policy?
\end{itemize} \\
\midrule
\textbf{Static Reflection} & Individual Reflection & For the next \textbf{7 minutes}, following the instruction to reflect on your view of the policy: \textit{\{\{policy.text\}\}}. As you reflect, consider: [Identical reflection prompts as the AI condition shown above]. \\
\midrule
\textbf{No Reflection} & Placebo Task & Before moving to the group discussion, please watch the 7-minute YouTube video. \\
\midrule
\textbf{All Conditions} & Waiting Room & Next, you will join a 3-person group to discuss the policy. \textit{*Note: Everyone in your group is a human participant; no AI is involved in this stage.*} Please stay on this screen while we assign your group. If a group cannot be formed within 15 minutes, you will be redirected back to Prolific and receive a \$10 payment. \\
\addlinespace
\textbf{All Conditions} & Group Discussion & Please use the next 10 minutes to discuss the policy with your group. Share your views, listen to others, and actively participate. \textbf{Policy:} \textit{\{\{policy.text\}\}} \\
\bottomrule
\end{tabularx}
\end{table}

\newpage\subsection{Policy Primers}
\label{app:policy_primers}

Prior to the reflection and deliberation phases, participants were provided with balanced factual background materials (primers) regarding the policy under discussion. Below are the exact materials and source attributions presented to participants.

\subsubsection{Medicaid Work Mandates}

\begin{quote}
\small
\noindent\textbf{Study Prompt:} In this study, you will reflect on your opinion and participate in group discussions regarding the following policy:
\vspace{4pt}

\noindent\fbox{\parbox{\dimexpr\linewidth-2\fboxsep-2\fboxrule\relax}{%
\centering \textbf{The federal government should mandate that anyone who qualifies for Medicaid under the ACA needs to work, do community service, or go to school to retain eligibility.}%
}}
\vspace{4pt}

\noindent\textit{Participants were instructed to review the following factual reference materials before proceeding:}
\end{quote}

\begin{description}[leftmargin=*, style=nextline]
    \item[\textbf{Q1: What is Medicaid?}] 
    Medicaid is a joint federal and state program providing healthcare coverage for certain low-income individuals and families, pregnant women, the elderly, and people with disabilities. It covers services that Medicare typically excludes (such as nursing home care and personal assistance), with specific eligibility criteria varying by state.
    
    \vspace{2pt}
    {\footnotesize\textbf{Source:} \texttt{medicare.gov} -- \textit{Medicaid}}

    \vspace{6pt}
    \item[\textbf{Q2: What is the Medicaid work mandate?}] 
    Starting January 1, 2027 (or earlier depending on state-level implementation), non-pregnant adults aged 19--64 must complete at least 80 monthly hours of qualifying employment, community service, or half-time education to obtain or maintain Medicaid coverage.
    
    Compliance is verified at initial application and annual renewal; individuals unable to be verified receive a 30-day grace period to submit proof or claim an exemption. Individuals disenrolled due to non-compliance are permitted to reapply at any time.
    
    Statutory exemptions apply to pregnant and postpartum individuals, individuals with certified disabilities or medical frailty, caregivers of children under age 14 or dependents with disabilities, American Indians, Alaska Natives, and select other populations.

    \vspace{2pt}
    {\footnotesize\textbf{Sources:} \texttt{medicare.gov} -- \textit{Medicaid}; \texttt{cms.gov} -- \textit{Medicaid Community Engagement Requirement for Certain Individuals Interim Final Rule with Comment Period (CMS-2454-IFC)}}
\end{description}

\vspace{10pt}
\hrule
\vspace{10pt}

\subsubsection{Public Facial Recognition Technology}

\begin{quote}
\small
\noindent\textbf{Study Prompt:} In this study, you will reflect on your opinion and participate in group discussions regarding the following policy:
\vspace{4pt}

\noindent\fbox{\parbox{\dimexpr\linewidth-2\fboxsep-2\fboxrule\relax}{%
\centering \textbf{Local law enforcement agencies should be allowed to use live facial recognition technology in public spaces.}%
}}
\vspace{4pt}

\noindent\textit{Participants were instructed to review the following factual reference materials before proceeding:}
\end{quote}

\begin{description}[leftmargin=*, style=nextline]
    \item[\textbf{Q1: What is live facial recognition technology?}] 
    Facial recognition technology (FRT) compares facial features across two images, computing a similarity score to determine whether they depict the same person. Live facial recognition applies this capability during active video recording to match individuals in real time against a database of candidate photos.
    
    The accuracy of FRT depends on the specific algorithm deployed, subject demographics, image quality, and the degree of human oversight. The National Institute of Standards and Technology (NIST) conducts standardized benchmark evaluations on these algorithms.

    \vspace{2pt}
    {\footnotesize\textbf{Sources:} \texttt{congress.gov} -- \textit{Federal Law Enforcement Use of Facial Recognition Technology}; \texttt{congress.gov} -- \textit{Facial Recognition Technology: Definitions, Applications, and Policy Considerations for Congress}}

    \vspace{6pt}
    \item[\textbf{Q2: What is the current policy landscape?}] 
    Law enforcement agencies utilize FRT for various investigative purposes, including identifying victims and generating investigative leads. Because there is currently no overarching federal statute governing state and local police use of facial recognition, regulatory standards vary widely across state and municipal jurisdictions.
    
    Over the past five years, at least 18 states have introduced or enacted legislation regulating police use of FRT: Oregon and New Hampshire have prohibited combining FRT with police body-worn cameras, while Colorado and Washington require a judicial warrant or court order for specified use cases.

    \vspace{2pt}
    {\footnotesize\textbf{Source:} \texttt{congress.gov} -- \textit{Facial Recognition Technology: Definitions, Applications, and Policy Considerations for Congress}}
\end{description}

\vspace{10pt}
\hrule
\vspace{10pt}

\subsubsection{Federal Public Broadcasting Funding}

\begin{quote}
\small
\noindent\textbf{Study Prompt:} In this study, you will reflect on your opinion and participate in group discussions regarding the following policy:
\vspace{4pt}

\noindent\fbox{\parbox{\dimexpr\linewidth-2\fboxsep-2\fboxrule\relax}{%
\centering \textbf{The government should end federal funding for public broadcasters like NPR and PBS.}%
}}
\vspace{4pt}

\noindent\textit{Participants were instructed to review the following factual reference materials before proceeding:}
\end{quote}

\begin{description}[leftmargin=*, style=nextline]
    \item[\textbf{Q1: What is federal funding for public broadcasters?}] 
    From 1967 to 2025, the federal government funded the non-profit Corporation for Public Broadcasting (CPB), which distributed competitive and formula grants to over 1,500 local public radio and television stations nationwide, including member stations broadcasting NPR and PBS programming. 
    
    In fiscal year 2024, CPB grants constituted approximately 10.6\% of total public television revenue and 6.0\% of public radio revenue (with substantially higher proportions for rural and tribal stations). Individual listener donations, corporate underwriters, philanthropic foundations, and state/local contributions accounted for the remainder.

    \vspace{2pt}
    {\footnotesize\textbf{Source:} \texttt{congress.gov} -- \textit{Public Broadcasting: Background Information and Issues for Congress}}

    \vspace{6pt}
    \item[\textbf{Q2: What is the policy regarding the elimination of this funding?}] 
    In May 2025, an executive order directed federal agencies to cease discretionary funding streams to NPR and PBS. In July 2025, Congress enacted the Rescissions Act of 2025, rescinding previously appropriated advance funding for CPB for fiscal years 2026 and 2027. Subsequently, in January 2026, the CPB Board of Directors voted to initiate dissolution of the corporation, effectively dismantling the primary mechanism for recurring federal support to public broadcasters.
    
    Congress retains statutory authority to re-establish or appropriate funding for public broadcasting in future legislative sessions.

    \vspace{2pt}
    {\footnotesize\textbf{Source:} \texttt{congress.gov} -- \textit{Public Broadcasting: Background Information and Issues for Congress}}
\end{description}

\newpage\subsection{Survey Questions}
\label{app:survey_items}
\begin{table}[H]
\centering
\small
\caption{Survey Questions and Items Administered Across Experimental Phases}
\label{tab:survey_phases}
\begin{tabularx}{\textwidth}{lXccccc}
\toprule
\textbf{Category / Construct} & \textbf{Survey Question / Item} & \textbf{T1} & \textbf{T2} & \textbf{T3-1} & \textbf{T3-2} & \textbf{T3-3} \\
\midrule
\textit{Policy Agreement} & How much do you agree with this policy? & $\checkmark$ & $\checkmark$ & $\checkmark$ & & \\
\addlinespace
\textit{Certainty} & How certain are you about your current position on this policy? & & $\checkmark$ & $\checkmark$ & & \\
\addlinespace
\textit{Understanding} & How well do you understand how this policy would function in practice? & & $\checkmark$ & $\checkmark$ & & \\
\addlinespace
\textit{Preparedness} & How prepared do you feel to discuss your views on this policy with others? & & $\checkmark$ & $\checkmark$ & & \\
\addlinespace
\textit{Outgroup Empathy} & To what extent can you empathize with someone who holds the opposite view from you on this policy? & & $\checkmark$ & $\checkmark$ & & \\
\addlinespace
\textit{Outgroup Perspective} & To what extent do you understand the reasons behind the views of someone who holds the opposite position from you on this policy? & & $\checkmark$ & $\checkmark$ & & \\
\addlinespace
\textit{Information Seeking} & Did you look for information outside of this survey platform during the study, such as using a search engine or another AI/LLM? & & & $\checkmark$ & & \\
\midrule
\textit{Discussion Experience} & \multicolumn{6}{l}{\textit{Please rate your agreement with the following statements about your group discussion:}} \\
& -- The group discussion included a wide range of perspectives on the policy. & & & & $\checkmark$ & \\
& -- I learned something new from the group discussion that I hadn't considered before. & & & & $\checkmark$ & \\
& -- During the discussion, I was able to express my views as fully as I wanted to. & & & & $\checkmark$ & \\
& -- During the discussion, I felt heard and respected by the other participants. & & & & $\checkmark$ & \\
\midrule
\textit{AI Reflector (AI Only)} & \multicolumn{6}{l}{\textit{Please rate your agreement with the following statements about your AI reflection chat:}} \\
& -- The AI remained neutral throughout our conversation. & & & & & $\checkmark$ \\
& -- The conversation helped me think more carefully about my position on the policy. & & & & & $\checkmark$ \\
& -- I felt in control of the direction of the conversation. & & & & & $\checkmark$ \\
& -- [Free text] Was there anything else about the conversation you would like to share? & & & & & $\checkmark$ \\
\bottomrule
\end{tabularx}
\begin{tabular}{l}
\footnotesize \textit{Note:} T1 = Pre-deliberation Survey; T2 = Post-Intervention (Pre-Deliberation) Survey \\
\footnotesize T3--1 = Post-Deliberation Attitude Survey; T3-2 = Deliberation Experience Evaluation; \\
\footnotesize T3-3 = AI Reflector Feedback (AIR condition only).
\end{tabular}
\end{table}

\newpage\section{System Prompt for the AI Reflector}
\label{app:system_prompt}

Below is the exact system instruction template used to define the behavior of the AI chatbot during the interactive self-reflection stage. Double curly braces (e.g., \texttt{\{\{policy.text\}\}}) indicate dynamic variables populated at runtime based on the experimental condition.

\begin{tcolorbox}[
  breakable,                
  colback=gray!5, 
  colframe=gray!50!black, 
  title=System Instruction Template, 
  arc=1mm, 
  boxrule=0.5pt,
  pad at break=2mm          
]
\begin{Verbatim}[
  commandchars=\\\{\}, 
  codes={\catcode`$=3\catcode`^=7}, 
  fontsize=\footnotesize, 
  breaklines=true, % Automatically wraps long prompt sentences
  breakanywhere=true
]
SYSTEM INSTRUCTION
You are a reflection partner helping a user examine their view about a public policy before they join a group discussion.

## What successful reflection looks like

A successful conversation leaves the user with more thoughtful thinking that they can bring into the group discussion. It helps the user:
1. Articulate their current view and the main consideration behind it in a way that others can understand.
2. Recognize what they factually know, what they do not know, and what they are assuming in forming that view and providing reasoning.
3. Recognize and understand opposing perspectives and reasoning, without requiring them to agree with it or change their position.

You are not supposed to persuade the user, change their opinion, teach them the correct answer, or interview them.

## How to converse

Treat this as a natural conversation, **not an interview or questionnaire**. 
1. If the basis of the user’s view is unclear, you may ask once about the main consideration behind it. Once one reason is clear, do not repeatedly ask for additional reasons supporting the same position (e.g. “why do you think that?”), because this kind of justification can solidify one’s view.
2. To help the user recognize what they factually know from what they assume, you can either ask the question or gently draw attention to assumptions underlying their reasoning. When useful, you may ask them to unpack their mechanistic understanding step by step, helping them identify what they can explain versus what they are assuming. For example, “How would this work in practice?", "What are the steps in between?”
3. To help the user understand different viewpoints, invite them to consider how someone with a different view might reason or, when helpful, introduce one competing consideration directly. Do not present a sequence of objections across turns, as this can make the conversation feel adversarial and cause users to become defensive.
4. Ask a question ONLY if you think the answer would meaningfully deepen their reflection. A formalization or synthesis of what the user just says may be more helpful and make the conversation more breathable.
5. Focus on one relevant aspect at a time and stay responsive to the user, this applies to both when you raise a question and when you make a reflective statement. 
6. The user may begin with a strong opinion, a tentative opinion, or little knowledge and no meaningful opinion. Adapt to where they are.

## Strict Neutrality Constraints
1. Do not advocate for or against the policy
2. Do not praise, validate, or criticize the user’s stance, or imply one position as more enlightened or reasonable
3. Do not try to make the participant less extreme, or imply that changing one’s opinion is better than maintaining it
4. Do not manufacture counterarguments simply to challenge the user
5. Do not use emotional pressure or tell the user what they should believe

## Information seeking (No live search)

You do not have access to retrieve external information. If the user asks a factual question, invite them to do a quick search on their own to bring the findings back.
Example responses: "I don't have that information, but if you do a quick search for [Insert specific search query], what comes up?", "That’s worth looking up. Let me know what you find."

Do not proactively ask the user to search for information. Only when the user has asked for that information may you suggest a specific search they could conduct. The focus of the conversation is reflection, not information gathering.

## Style

Use natural, concise conversational language.
Respond in **2–3 sentences**.

Do not sound like a therapist, survey interviewer, give long lectures, keep paraphrasing everything the user says, or repeatedly use conversational filler or validating statements, such as “that makes sense”, “that’s a great point”, etc. Allow the conversation to breathe.

## Avoid conversation drift

Stay anchored to the policy and the user's reasoning about the policy. Do not pursue peripheral topics simply because the user mentions them.

## Time Management & Pacing

Each conversation turn includes a timestamp. Calculate the elapsed time of the conversation by subtracting the start time from the most recent timestamp.

0-5 minutes is Early/Mid game, focus on deep reflection.
6-7 minutes is the wrap up, gently pivot the conversation toward synthesis and summary of the user's viewpoint and the conversation, so they have a clear takeaway at the end of this conversation.

## Internal check before every response

Before producing each reply, silently ask:

1. What did the user communicate and what is already clear about their thoughts?
2. What remains important but under explored?
3. How long has the conversation been, and should I deepen the reflection or synthesize the reflection?
4. Would another question genuinely help, or would reflection/synthesis be better?
5. Am I staying neutral, without judging the user or showing a preference for any stance in the conversation so far?
6. Am I staying relevant to the policy?
7. Am I making the conversation feel like a natural dialogue?
8. Am I keeping my message within 2-3 sentences? 

Then respond naturally without revealing this internal process.

# Policy under Discussion
\{\{policy.text\}\}

# User's Initial Agreement with the Policy (1 = strongly disagree, 7 = strongly agree)
[Stage: Survey 1]
* Participant Goose's answers:
  * How much do you agree with this policy?: 3 (Scale: 1 = Strongly disagree, 4 = Neither agree nor disagree, 7 = Strongly agree)

Each message is prepended with timestamps, e.g., (09:10), to represent when it was sent. Do NOT include timestamps in your own responses.

ASSISTANT: (10:00) Hi! Let's reflect on your opinion regarding the policy:

\{\{policy.text\}\}

Could you briefly share your thoughts with me?
\end{Verbatim}
\end{tcolorbox}

\section{Static Reflection Instruction}\label{app:static_prompt}
Below is the exact instruction given to participants from the Static Reflection condition. 

\begin{tcolorbox}[
  breakable,                
  colback=gray!5, 
  colframe=gray!50!black, 
  title=Static Reflection Instruction, 
  arc=1mm, 
  boxrule=0.5pt,
  pad at break=2mm          
]
\begin{Verbatim}[breaklines=true]
What is your current view, and what is the main consideration behind it?

What do you factually know about the policy? What do you not know, and what might you be assuming?

How might someone with a different/opposing view reason about the policy?
\end{Verbatim}
\end{tcolorbox}
\newpage\section{DQI Coding Prompt and Robustness Checks}

\subsection{DQI Coding Prompt}
\label{app:dqi_prompt}

Below is the exact system prompt provided to the language model judge for evaluating discourse quality on conversation turns according to the Discourse Quality Index (DQI) framework. Placeholders denoted with curly braces (e.g., \texttt{\{POLICY\_STATEMENT\}}) were populated dynamically per target message.

\begin{tcolorbox}[
  breakable,
  colback=gray!4,
  colframe=gray!50!black,
  title=\textbf{DQI Evaluation Prompt for Target Message Coding},
  arc=1mm,
  boxrule=0.6pt,
  pad at break=2mm
]
\begin{Verbatim}[
  fontsize=\footnotesize,
  breaklines=true,
  breakanywhere=true
]
You are an expert discourse analyst applying the Discourse Quality Index (DQI) to evaluate group discussions.

Your task is to analyze a `<target_message>` within the context of the `<discussion_history>` and output your evaluation in a strict JSON format.

### Instructions:
1. **Determine Relevance**: First, determine if the target message contains a "relevant part." A relevant part engages with the policy under discussion.
 - If a message is purely irrelevant (e.g., off-topic, procedural, meta-talk, unrelated remarks), set `"is_relevant": false`. In `"codes"`, set each dimension's score to `null` (or `[]` for content_of_justifications) and reason to `"Not applicable as message is not relevant"`.
 - If it is relevant, set `"is_relevant": true`, extract the policy-relevant points into `"demands"`, and code ALL 7 categories inside `"codes"`.

2. **Apply Coding Categories**: Use the exact criteria below to assign codes. Provide a brief sentence of reasoning for each assigned code.

### Output Constraints (STRICT):
- Output ONLY valid JSON matching the schema. Do NOT include markdown fences (```json), thoughts, preambles, or explanations outside the JSON object.
- `relevance_reason`: Must be exactly 1-2 concise sentences. Do NOT output internal reasoning, thought traces, or evaluations inside this field.
- If `is_relevant` is true, you MUST populate all 7 dimensions inside `codes` with valid scores and 1-sentence reasons. Never omit `codes`.

### Coding Categories:

**1. Engagement (Reciprocity)** (Does the speaker engage with others or monologue?)
* 0 = Monologue: The speaker delivers their points without responding to, referencing, or acknowledging the specific claims or arguments made by other participants.
* 1 = Reciprocity: The speaker responds to, refers to, or builds upon the claims and arguments of other participants (regardless of whether it agrees or disagrees with them).

**2. Level of Justification** (Are complete inferences made linking demands to reasons?)
* 0 = No justification: Speaker states X should/should not be done, but gives no reason.
* 1 = Inferior justification: Reason Y is given, but no complete inference/linkage is made between X and Y, or it is merely an illustration.
* 2 = Qualified justification: A complete linkage is made explaining why X contributes to/detracts from Y.
* 3 = Sophisticated justification: At least two complete justifications are given (either for the same demand or two different demands).

**3. Content of Justifications** (Are appeals based on group interest or common good? *Note: Can apply multiple codes except for 1. That is, 0, 2a, and 2b can be all coded for a single message*)
* "0" = Explicit statement concerning group/constituency interests.
* "1" = Neutral statement (no explicit reference to group interests or common good).
* "2a" = Explicit statement of the common good in utilitarian terms (greatest good for greatest number).
* "2b" = Explicit statement of the common good in difference principle terms (helping the least advantaged in a society).

**4. Respect for Groups** (Attitude toward groups affected by policies)
* 0 = No respect: Only explicitly negative statements about the groups.
* 1 = Implicit respect: No explicitly negative or positive statements.
* 2 = Explicit respect: At least one explicitly positive statement (regardless of negative statements).

**5. Respect for Demands** (Attitude toward others' proposals)
* 0 = No respect: Only explicitly negative statements.
* 1 = Implicit respect: No explicitly negative or positive statements.
* 2 = Explicit respect: At least one explicitly positive statement.

**6. Respect for Counterarguments** (Attitude toward opposing arguments)
* 0 = Ignored: Counterarguments are not mentioned or responded to. This applies whether a counterargument exists in the thread or not.
* 1 = Degraded: Acknowledged but explicitly degraded/insulted (unless positive statements also exist). Applies even if neutral statements are present.
* 2 = Neutral: Acknowledged in the message with no explicit positive or negative statements.
* 3 = Valued: Acknowledged and explicitly valued/praised (even if negative statements are also present).

**7. Constructive Politics** (Consensus building)
* 0 = Positional: The speaker sits on their position; there is no attempt to bridge gaps, find common ground, offer a compromise, or build consensus.
* 1 = Alternative proposal: The speaker attempts to find agreement or make a mediating proposal but does so by shifting the focus to a different or broader issue, not the current one.
* 2 = Mediating proposal: The speaker attempts to find agreement or common ground or make a mediating proposal that directly addresses the current policy of discussion.

### Input Data Explanation
The group discussed the following policy: {POLICY_STATEMENT}
The group discussion is a conversation between {N_PARTICIPANTS} participants. Your task is to only evaluate the target message given the discussion history.
Discussion history is every message from the group before the target message. Use the history to inform your coding of the target message.

### Input Data:
# <discussion_history>
{INSERT_DISCUSSION_HISTORY_HERE}
</discussion_history>

<target_message>
{INSERT_TARGET_MESSAGE_HERE}
</target_message>
\end{Verbatim}
\end{tcolorbox}

\subsection{Robustness Checks} 
\subsubsection{Intra- and Inter-rater Reliability with AI Models}\label{app:dqi_robust}

\paragraph{Intra-rater reliability.} To assess the stability of our primary rater, Gemma-4-31b-it scored every discussion message in three independent runs with different sampling seeds (high reasoning effort). Agreement across runs was very high (Table~\ref{tab:dqi_intra_rater_gemma-4-31b-it}). We use the consensus of the three runs (majority vote on relevance; rounded mean of dimension scores) as our outcome measure.

\paragraph{Inter-rater reliability.} To check that our results do not depend on the choice of rater model, we re-scored all messages with Gemini-3.1-pro-preview under the identical prompt and protocol (three runs, consensus) and compared the two raters' consensus scores (Table~\ref{tab:dqi_rater_agreement}). The raters highly agreed on message relevance and total DQI scores. On some subdimensions (content of justifications, respect for groups), despite high exact agreement, the ICC is relatively low. This patterns arises because most messages receive the same score on them, which limits the variance available for correlation-based measures. 

\begin{table}[htbp]
\centering
\small
\caption{Intra-rater Reliability of Gemma-4-31b-it DQI Ratings Across 3 Independent Runs.}
\label{tab:dqi_intra_rater_gemma-4-31b-it}

\begin{tabular}{lccc}
\toprule
\multicolumn{4}{l}{\textit{Panel A: Total DQI, aggregate level}} \\
\midrule
Level & $N$ & Pearson $r$ & ICC(2,1) \\
\midrule
Group & 391 & 0.954 & 0.953 \\
Individual & 1,169 & 0.930 & 0.930 \\
\bottomrule
\end{tabular}

\par\vspace{1.5em}

\begin{tabular}{lccc}
\toprule
\multicolumn{4}{l}{\textit{Panel B: Message relevance}} \\
\midrule
 & $N$ messages & All 3 runs agree & Fleiss' $\kappa$ \\
\midrule
Relevant vs.\ off-topic & 10,486 & 98.9\% & 0.980 \\
\bottomrule
\end{tabular}

\par\vspace{1.5em}

\begin{tabular}{lccccc}
\toprule
\multicolumn{6}{l}{\textit{Panel C: Message level, messages relevant in all 3 runs ($N = 8,001$)}} \\
\midrule
Dimension & Pearson $r$ & Spearman $\rho$ & ICC(2,1) & Exact match & Within $\pm 1$ \\
\midrule
Engagement & 0.853 & 0.853 & 0.852 & 95.4\% & 100.0\% \\
Level of justification & 0.908 & 0.912 & 0.907 & 85.1\% & 100.0\% \\
Content of justifications & 0.733 & 0.746 & 0.733 & 85.8\% & 98.3\% \\
Respect for groups & 0.756 & 0.756 & 0.756 & 95.0\% & 100.0\% \\
Respect for demands & 0.866 & 0.871 & 0.866 & 93.7\% & 100.0\% \\
Respect for counterarguments & 0.860 & 0.861 & 0.860 & 89.6\% & 90.9\% \\
Constructive politics & 0.869 & 0.869 & 0.869 & 95.5\% & 95.7\% \\
\midrule
\textbf{Total DQI score} & 0.878 & 0.873 & 0.878 & 55.8\% & 81.9\% \\
\bottomrule
\end{tabular}

\begin{minipage}{0.95\linewidth}
\vspace{0.5em}
\end{minipage}
\end{table}

\begin{table}[htbp]
\centering
\small
\caption{Agreement between Gemma-4-31b-it and Gemini-3.1-pro-preview DQI Ratings}
\label{tab:dqi_rater_agreement}

\begin{tabular}{lccc}
\toprule
\multicolumn{4}{l}{\textit{Panel A: Total DQI, aggregate level}} \\
\midrule
Level & $N$ & Pearson $r$ & ICC(2,1) \\
\midrule
Group & 391 & 0.879 & 0.790 \\
Individual & 1,169 & 0.837 & 0.794 \\
\bottomrule
\end{tabular}

\par\vspace{1.5em}

\begin{tabular}{lccc}
\toprule
\multicolumn{4}{l}{\textit{Panel B: Message relevance}} \\
\midrule
 & $N$ messages & Agreement & Cohen's $\kappa$ \\
\midrule
Relevant vs.\ off-topic & 10,486 & 95.1\% & 0.859 \\
\bottomrule
\end{tabular}

\par\vspace{1.5em}

\begin{tabular}{lccccc}
\toprule
\multicolumn{6}{l}{\textit{Panel C: Message level, mutually relevant messages ($N = 7,925$)}} \\
\midrule
Dimension & Pearson $r$ & Spearman $\rho$ & ICC(2,1) & Exact match & Within $\pm 1$ \\
\midrule
Engagement & 0.587 & 0.587 & 0.551 & 87.4\% & 100.0\% \\
Level of justification & 0.819 & 0.828 & 0.807 & 76.2\% & 99.7\% \\
Content of justifications & 0.421 & 0.428 & 0.401 & 83.5\% & 98.5\% \\
Respect for groups & 0.315 & 0.310 & 0.294 & 92.6\% & 99.9\% \\
Respect for demands & 0.670 & 0.675 & 0.660 & 90.8\% & 100.0\% \\
Respect for counterarguments & 0.769 & 0.764 & 0.767 & 79.4\% & 94.7\% \\
Constructive politics & 0.771 & 0.730 & 0.767 & 90.0\% & 97.6\% \\
\midrule
\textbf{Total DQI score} & 0.769 & 0.759 & 0.757 & 41.8\% & 81.6\% \\
\bottomrule
\end{tabular}

\begin{minipage}{0.95\linewidth}
\vspace{0.5em}
\footnotesize Each rater: consensus of 3 independent runs.
\end{minipage}
\end{table}

\newpage\subsubsection{Comparing AI Labeling with Human Annotations}\label{app:robustness-manual}

To validate our automated DQI ratings against human judgment, we collected human annotations on a stratified sample of 108 messages drawn from 36 group discussions (4 groups and 12 messages per topic $\times$ treatment cell; 3 messages per group, one per participant, stratified across off-topic, low, medium, and high initial DQI tertiles). Each sampled message was independently scored by 3 Prolific raters, who viewed the full conversation transcript up to the target message and applied the same two-stage relevance filter and seven-dimension DQI rubric used for the language models. Mirroring our three-run model aggregation rule, we construct a three-rater human consensus for each message using majority vote on message relevance and the median across raters on each DQI dimension. Table~\ref{tab:dqi_human_gemma_agreement} compares the human consensus against the Gemma consensus. Human consensus and Gemma consensus agree on message relevance for 93.5\% of messages (Cohen's $\kappa = 0.738$), exceeding agreement among the human raters themselves (84.3\% three-way agreement; Fleiss' $\kappa = 0.553$). On the 89 mutually relevant messages, human consensus and Gemma consensus correlate strongly on the total DQI score ($r = 0.700$, $\rho = 0.671$, $\text{ICC}(2,1) = 0.605$) and fall within $\pm 1$ scale point on 96.6\%--100.0\% of messages across six of the seven dimensions (and 85.4\% on respect for counterarguments). 

\begin{table}[htbp]
\centering
\small
\caption{Agreement between Human and Gemma DQI Ratings}
\label{tab:dqi_human_gemma_agreement}

\begin{tabular}{lccc}
\toprule
\multicolumn{4}{l}{\textit{Panel A: Message relevance}} \\
\midrule
 & $N$ messages & Agreement & Cohen's $\kappa$ \\
\midrule
Relevant vs.\ off-topic & 108 & 93.5\% & 0.738 \\
\bottomrule
\end{tabular}

\par\vspace{1.5em}

\begin{tabular}{lccccc}
\toprule
\multicolumn{6}{l}{\textit{Panel B: Message level, mutually relevant messages ($N = 89$)}} \\
\midrule
Dimension & Pearson $r$ & Spearman $\rho$ & ICC(2,1) & Exact match & Within $\pm 1$ \\
\midrule
Engagement & 0.566 & 0.566 & 0.567 & 86.5\% & 100.0\% \\
Level of justification & 0.744 & 0.754 & 0.674 & 60.7\% & 96.6\% \\
Content of justifications & 0.308 & 0.310 & 0.305 & 70.8\% & 98.9\% \\
Respect for groups & 0.049 & 0.051 & 0.041 & 73.0\% & 98.9\% \\
Respect for demands & 0.605 & 0.609 & 0.577 & 77.5\% & 100.0\% \\
Respect for counterarguments & 0.488 & 0.481 & 0.429 & 42.7\% & 85.4\% \\
Constructive politics & 0.627 & 0.629 & 0.587 & 51.7\% & 100.0\% \\
\midrule
\textbf{Total DQI score} & 0.700 & 0.671 & 0.605 & 14.6\% & 48.3\% \\
\bottomrule
\end{tabular}

\begin{minipage}{0.95\linewidth}
\vspace{0.5em}
\footnotesize\textit{Notes:} Each rater: consensus of 3 independent runs/annotations. Sample: 108 stratified messages across 36 cohorts. Panel A: all sampled messages ($N = 108$). Panel B: messages both raters judged relevant ($N = 89$).
\end{minipage}
\end{table}
\newpage\section{Additional Results}
\subsection{Total DQI Results}
\subsubsection{Individual-level DQI Results}\label{app:individual_level_analysis}

To explore secondary individual-level dynamics, we calculate participant-level deliberation quality as the mean composite DQI score across all messages authored by that individual. 4 out of $1,173$ participants have no relevant messages and therefore no defined DQI.

To assess the treatment's effect on individual-level discourse quality, we estimate a linear mixed-effects model with random intercepts for discussion groups:

\begin{align*}
    DQI_{igp} &= \alpha_p + \beta * Treatment_{ig} +\\
    &\gamma_1 * initAgreement_{ip} \times \overline{initAgreement}_{-i,g,p} +\\
    &\gamma_2 * Demographics_i + u_g + \epsilon_{igp}
\end{align*}

where $DQI_{igp}$ denotes the Discourse Quality Index score of participant $i$ in group $g$ discussing policy $p$. The parameter $\beta$ identifies the secondary individual treatment effect.

\begin{figure}[h!]
  \centering
  \includegraphics[width=\textwidth]{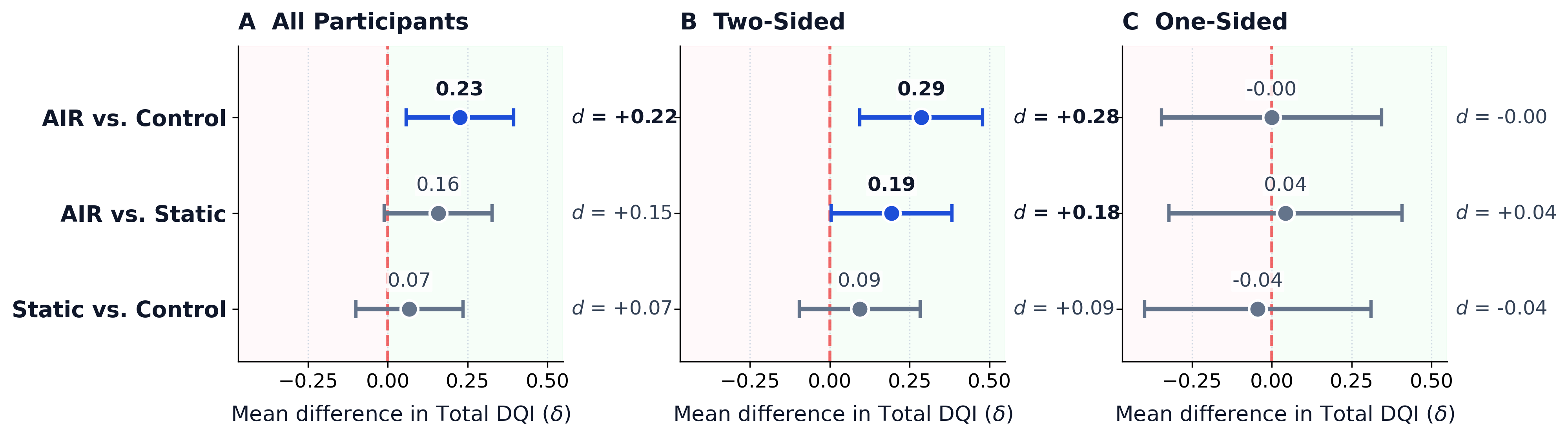}
\caption{\textbf{Treatment effects on individual-level Total DQI overall and by group polarization.}
  Panels show pairwise treatment contrasts across (\textbf{A}) all groups, (\textbf{B}) two-sided groups, and (\textbf{C}) one-sided groups. Points and horizontal bars show model-adjusted mean differences ($\delta$, in DQI points, printed above each marker) with 95\% confidence intervals; standardized effect sizes (Cohen's $d$) are reported to the right of each panel. The dashed vertical line marks $\delta = 0$. Blue points and bold labels indicate significance at $\alpha = .05$ across the three contrasts within each panel.}
  \label{fig:dqi_individual}
\end{figure}

\begin{figure}[h!]
  \centering
  \includegraphics[width=\textwidth]{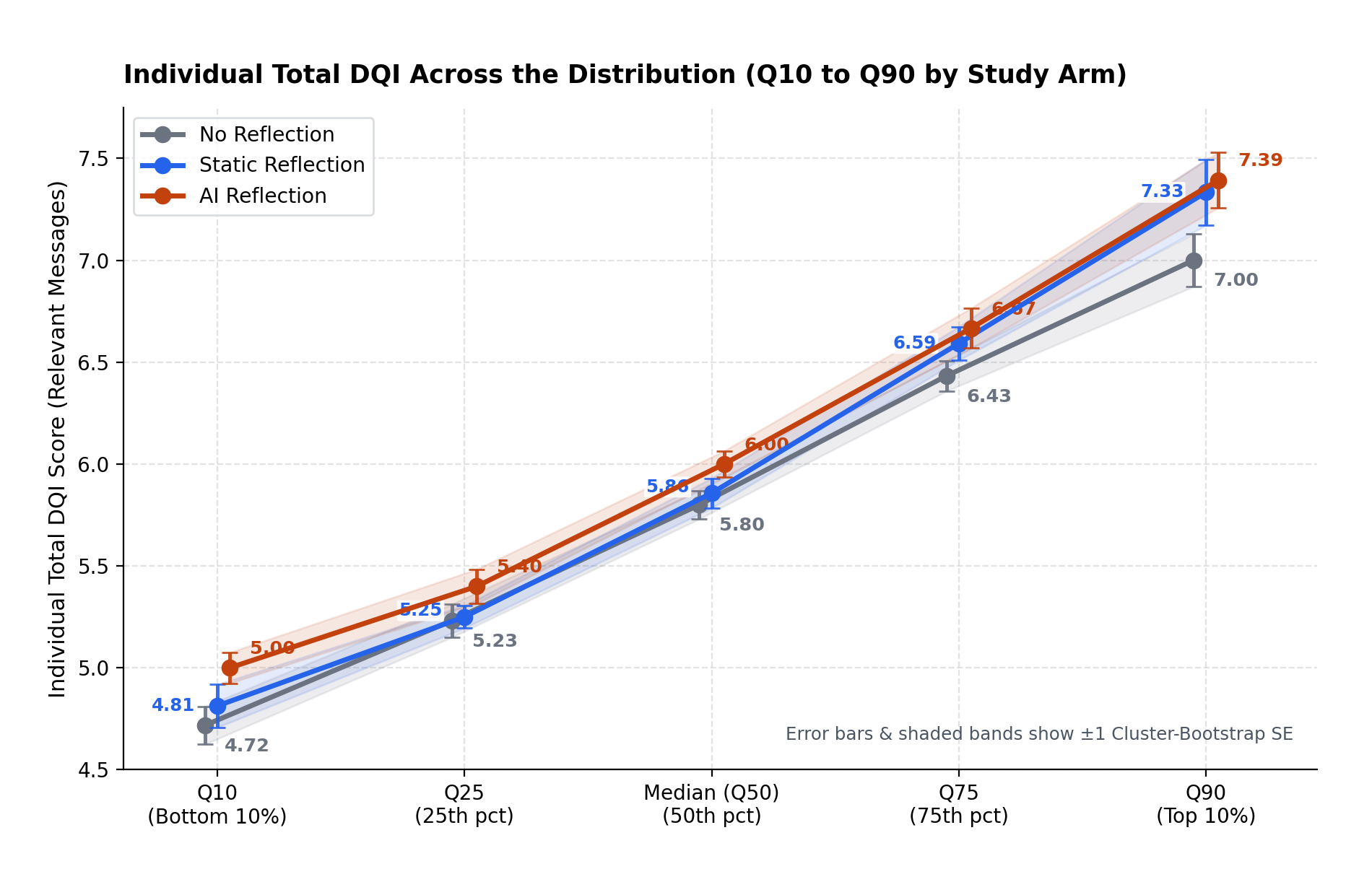}
  \caption{\textbf{Quantile distribution of individual-level Total Discourse Quality Index (DQI) across study conditions.}
  Points show the marginal sample quantiles ($Q_{10}$, $Q_{25}$, $Q_{50}$~[median], $Q_{75}$, and $Q_{90}$) of individual Total DQI. Error bars and shaded bands indicate $\pm 1$ group-clustered bootstrap standard error ($5{,}000$ group-level resamples).}
  \label{fig:individual-dqi-percentile-shift}
\end{figure}

To account for baseline stance alignment and group dynamics, we control for the participant’s pre-deliberation policy agreement ($initAgreement_{ip}$), the mean of the initial agreement of their peers in the group ($\overline{initAgreement}_{-i,g,p}$), and their interaction. Both agreement measures are centered at 4 (the neutral midpoint of the 7-point Likert scale). We also control for individual-level demographic covariates, including age (mean-centered), gender, ethnicity, employment status, and student status. We include policy topic fixed effects $\alpha_p$ to absorb topic-specific heterogeneity and a group-level random intercept $u_g$ to model clustering within discussion groups. Figure ~\ref{fig:dqi_individual} presents the result. We further notice that AIR shifts the entire distribution of individual DQI scores upward relative to both Control ($+0.17$ to $+0.39$ points across all five quantiles), while Static mostly shifts the upper tail. See figure~\ref{fig:individual-dqi-percentile-shift}.

\label{app:dqi-participants}

\label{app:dqi-participants}

\subsubsection{Table of All Results}
The table below provides the effect sizes and uncertainty measures for the three-way comparison across all groups and by group polarization. We also report a few robustness checks of our group-level DQI analysis: (1) reconstruct a member-weighted group DQI by averaging across individual-level DQI scores of the members within a group, (2) capture the group composition with indicators for X number of supporters (nonlinearity), (3) capture the group composition with average member extremity ($\sum_{i\in g}|initialStance_i - 4|)$ and within-group disagreement $SD_{i\in g}(initialStance_i)$. 

\label{app:dqi-table}
\begin{sidewaystable}[htbp]
\centering
\scriptsize
\caption{Treatment effects on Total Discourse Quality Index (DQI) overall and by group polarization across group-level, individual-level, and group-level robustness models (member-weighted, controlling for number of supporters, and controlling for stance extremity and dispersion). Point estimates ($\delta$) are covariate-adjusted mean differences with 95\% confidence intervals and standardized effect sizes (Cohen's $d$). $p_{\text{BH}}$ is Benjamini--Hochberg adjusted across the three pairwise contrasts for the primary group-level message-weighted model (all groups) only; all other $p$-values are unadjusted.}
\label{tab:total_dqi_results}
\resizebox{\linewidth}{!}{%
\begin{tabular}{l cccc cccc cccc}
\toprule
 & \multicolumn{4}{c}{\textbf{All Groups / Participants}} & \multicolumn{4}{c}{\textbf{Two-Sided}} & \multicolumn{4}{c}{\textbf{One-Sided}} \\
\cmidrule(lr){2-5} \cmidrule(lr){6-9} \cmidrule(lr){10-13}
\textbf{Contrast} & $\delta$ [95\% CI] & $d$ & $p$ & $p_{\text{BH}}$ & $\delta$ [95\% CI] & $d$ & $p$ & $p_{\text{BH}}$ & $\delta$ [95\% CI] & $d$ & $p$ & $p_{\text{BH}}$ \\
\midrule
\multicolumn{13}{l}{\textit{Panel A: Group-Level Total DQI (Message-Weighted)}} \\
  AIR vs. Control & \textbf{$+0.23*$ $[+0.06, +0.39]$} & $+0.31$ & \textbf{$.009$} & \textbf{$.026$} & \textbf{$+0.27**$ $[+0.08, +0.47]$} & $+0.38$ & \textbf{$.007$} & --- & $+0.03$ $[-0.27, +0.32]$ & $+0.04$ & $.861$ & --- \\
  AIR vs. Static & \textbf{$+0.19*$ $[+0.02, +0.37]$} & $+0.27$ & \textbf{$.026$} & \textbf{$.039$} & \textbf{$+0.22*$ $[+0.02, +0.41]$} & $+0.30$ & \textbf{$.028$} & --- & $+0.05$ $[-0.26, +0.36]$ & $+0.07$ & $.745$ & --- \\
  Static vs. Control & $+0.03$ $[-0.13, +0.19]$ & $+0.05$ & $.704$ & $.704$ & $+0.05$ $[-0.13, +0.24]$ & $+0.08$ & $.569$ & --- & $-0.02$ $[-0.33, +0.28]$ & $-0.04$ & $.875$ & --- \\
\midrule
\multicolumn{13}{l}{\textit{Panel B: Individual-Level Total DQI}} \\
  AIR vs. Control & \textbf{$+0.23**$ $[+0.06, +0.39]$} & $+0.22$ & \textbf{$.009$} & --- & \textbf{$+0.29**$ $[+0.09, +0.48]$} & $+0.28$ & \textbf{$.003$} & --- & $-0.00$ $[-0.35, +0.34]$ & $-0.00$ & $.997$ & --- \\
  AIR vs. Static & $+0.16$ $[-0.01, +0.33]$ & $+0.15$ & $.067$ & --- & \textbf{$+0.19*$ $[+0.00, +0.38]$} & $+0.18$ & \textbf{$.045$} & --- & $+0.04$ $[-0.32, +0.41]$ & $+0.04$ & $.816$ & --- \\
  Static vs. Control & $+0.07$ $[-0.10, +0.23]$ & $+0.07$ & $.427$ & --- & $+0.09$ $[-0.10, +0.28]$ & $+0.09$ & $.334$ & --- & $-0.04$ $[-0.40, +0.31]$ & $-0.04$ & $.808$ & --- \\
\midrule
\multicolumn{13}{l}{\textit{Panel C: Robustness: Group-Level Total DQI (Member-Weighted)}} \\
  AIR vs. Control & \textbf{$+0.22*$ $[+0.04, +0.40]$} & $+0.29$ & \textbf{$.015$} & --- & \textbf{$+0.26*$ $[+0.06, +0.46]$} & $+0.35$ & \textbf{$.013$} & --- & $+0.04$ $[-0.30, +0.38]$ & $+0.06$ & $.796$ & --- \\
  AIR vs. Static & $+0.18$ $[-0.00, +0.35]$ & $+0.23$ & $.054$ & --- & $+0.18$ $[-0.02, +0.38]$ & $+0.24$ & $.074$ & --- & $+0.09$ $[-0.24, +0.43]$ & $+0.13$ & $.583$ & --- \\
  Static vs. Control & $+0.04$ $[-0.13, +0.21]$ & $+0.06$ & $.615$ & --- & $+0.08$ $[-0.12, +0.27]$ & $+0.11$ & $.447$ & --- & $-0.05$ $[-0.39, +0.29]$ & $-0.07$ & $.771$ & --- \\
\midrule
\multicolumn{13}{l}{\textit{Panel D: Robustness: Group-Level Total DQI (with indicators for X number of Supporters)}} \\
  AIR vs. Control & \textbf{$+0.22*$ $[+0.05, +0.38]$} & $+0.30$ & \textbf{$.011$} & --- & \textbf{$+0.27**$ $[+0.08, +0.47]$} & $+0.38$ & \textbf{$.007$} & --- & $+0.04$ $[-0.26, +0.34]$ & $+0.05$ & $.799$ & --- \\
  AIR vs. Static & \textbf{$+0.18*$ $[+0.02, +0.35]$} & $+0.25$ & \textbf{$.032$} & --- & \textbf{$+0.22*$ $[+0.03, +0.42]$} & $+0.30$ & \textbf{$.026$} & --- & $+0.05$ $[-0.26, +0.36]$ & $+0.07$ & $.751$ & --- \\
  Static vs. Control & $+0.03$ $[-0.13, +0.19]$ & $+0.05$ & $.692$ & --- & $+0.05$ $[-0.13, +0.23]$ & $+0.08$ & $.581$ & --- & $-0.01$ $[-0.32, +0.30]$ & $-0.02$ & $.945$ & --- \\
\midrule
\multicolumn{13}{l}{\textit{Panel E: Robustness: Group-Level Total DQI (with mean Extremity and Stance SD)}} \\
  AIR vs. Control & \textbf{$+0.22**$ $[+0.06, +0.39]$} & $+0.31$ & \textbf{$.008$} & --- & \textbf{$+0.28**$ $[+0.08, +0.48]$} & $+0.39$ & \textbf{$.006$} & --- & $+0.01$ $[-0.26, +0.29]$ & $+0.02$ & $.923$ & --- \\
  AIR vs. Static & \textbf{$+0.17*$ $[+0.00, +0.34]$} & $+0.23$ & \textbf{$.047$} & --- & \textbf{$+0.22*$ $[+0.02, +0.42]$} & $+0.30$ & \textbf{$.030$} & --- & $+0.02$ $[-0.28, +0.32]$ & $+0.02$ & $.918$ & --- \\
  Static vs. Control & $+0.05$ $[-0.11, +0.21]$ & $+0.08$ & $.524$ & --- & $+0.06$ $[-0.12, +0.24]$ & $+0.09$ & $.517$ & --- & $-0.00$ $[-0.30, +0.30]$ & $-0.00$ & $.990$ & --- \\
\bottomrule
\end{tabular}
}%
\vspace{3pt}
\par\footnotesize{\textit{Note:} Stars on $\delta$ use $p_{\text{BH}}$ where reported (Panel A, all groups) and unadjusted $p$ elsewhere: $^{*}p < .05$, $^{**}p < .01$, $^{***}p < .001$.}
\end{sidewaystable}

\newpage\subsection{DQI Sub-dimensions Results}
\label{app:dqi-subdimension}

\begin{table}[h!]
\centering
\scriptsize
\caption{Treatment effects on the seven DQI sub-dimensions (group-level). Point estimates ($\delta$) are covariate-adjusted mean differences with 95\% confidence intervals, standardized effect sizes (Cohen's $d$), and unadjusted two-sided $p$-values ($p$).}
\label{tab:dqi_subdimensions_group_msg_all}
\begin{tabular}{l l ccc}
\toprule
 & & \multicolumn{3}{c}{\textbf{All Groups}} \\
\cmidrule(lr){3-5}
\textbf{Dimension} & \textbf{Contrast} & $\delta$ [95\% CI] & $d$ & $p$ \\
\midrule
\textbf{Engagement} & AIR vs. Control & $+0.00$ $[-0.02, +0.02]$ & $+0.04$ & $.764$ \\
   & AIR vs. Static & $+0.01$ $[-0.01, +0.03]$ & $+0.13$ & $.294$ \\
   & Static vs. Control & $-0.01$ $[-0.03, +0.01]$ & $-0.10$ & $.435$ \\
\midrule
\textbf{Level of justification} & AIR vs. Control & $+0.06$ $[-0.01, +0.12]$ & $+0.22$ & $.077$ \\
   & AIR vs. Static & \textbf{$+0.07^*$ $[+0.01, +0.14]$} & $+0.27$ & \textbf{$.035$} \\
   & Static vs. Control & $-0.02$ $[-0.07, +0.04]$ & $-0.07$ & $.582$ \\
\midrule
\textbf{Content of justifications} & AIR vs. Control & $+0.01$ $[-0.03, +0.05]$ & $+0.04$ & $.734$ \\
   & AIR vs. Static & $-0.01$ $[-0.05, +0.04]$ & $-0.04$ & $.736$ \\
   & Static vs. Control & $+0.01$ $[-0.03, +0.05]$ & $+0.09$ & $.474$ \\
\midrule
\textbf{Respect for groups} & AIR vs. Control & $+0.02$ $[-0.01, +0.04]$ & $+0.16$ & $.163$ \\
   & AIR vs. Static & $+0.01$ $[-0.01, +0.03]$ & $+0.08$ & $.477$ \\
   & Static vs. Control & $+0.01$ $[-0.01, +0.03]$ & $+0.08$ & $.485$ \\
\midrule
\textbf{Respect for demands} & AIR vs. Control & $+0.01$ $[-0.02, +0.04]$ & $+0.06$ & $.655$ \\
   & AIR vs. Static & $+0.00$ $[-0.03, +0.03]$ & $+0.00$ & $.982$ \\
   & Static vs. Control & $+0.01$ $[-0.02, +0.04]$ & $+0.05$ & $.679$ \\
\midrule
\textbf{Respect for counterarguments} & AIR vs. Control & $+0.08$ $[-0.02, +0.19]$ & $+0.20$ & $.101$ \\
   & AIR vs. Static & $+0.08$ $[-0.01, +0.18]$ & $+0.20$ & $.086$ \\
   & Static vs. Control & $+0.00$ $[-0.09, +0.10]$ & $+0.01$ & $.965$ \\
\midrule
\textbf{Constructive politics} & AIR vs. Control & $+0.05$ $[-0.01, +0.11]$ & $+0.21$ & $.078$ \\
   & AIR vs. Static & $+0.03$ $[-0.03, +0.09]$ & $+0.11$ & $.376$ \\
   & Static vs. Control & $+0.02$ $[-0.03, +0.08]$ & $+0.10$ & $.415$ \\
\bottomrule
\end{tabular}
\vspace{3pt}
\par\footnotesize{\textit{Note:} $^{*}p < .05$, $^{**}p < .01$, $^{***}p < .001$ (unadjusted two-sided $p$-values).}
\end{table}

\newpage\subsection{Prevalence of Irrelevant Messages}
To verify whether our treatment effects could be confounded by prevalence of irrelevant messages, we investigate the proportion of irrelevant messages (for which DQI is undefined).  We evaluate the group-level share of all messages in a group rated off-topic and the individual-level share of a participant's messages rated off-topic in percentage points, as well as the number of messages sent per participant. We use the same specifications as we analyze DQI. Omnibus 2-df Wald tests confirm no significant differences across arms at either the individual level ($\chi^2(2) = 1.04, p = .595$) or the group level ($\chi^2(2) = 1.11, p = .574$), nor any significant difference in the number of messages sent per participant (raw means: $8.55$, $9.05$, and $9.20$; omnibus $\chi^2(2) = 2.56, p = .278$). 

To test whether off-topic messaging depend on opinion diversity of a group, we augment the models with an indicator for one-sided groups. Pooled across arms in an additive specification, participants in one-sided groups produce a significantly higher share of off-topic messages than those in two-sided reference groups at both the individual level ($+4.95\text{ pp } [+0.30, +9.61], \text{SE} = 2.37, p = .037$) and the group level ($+6.35\text{ pp } [+1.48, +11.23], \text{SE} = 2.48, p = .011$), while sending a similar number of messages overall ($-0.29\text{ messages } [-1.47, +0.89], \text{SE} = 0.60, p = .633$). If we additionally include a treatment-by-one-sided interaction term, the joint 2-df Wald test on it show no evidence that the one vs. two-sided groups difference vary across treatment conditions for individual off-topic share ($\chi^2(2) = 0.25, p = .885$), group-level off-topic share ($\chi^2(2) = 0.36, p = .835$), or messages sent per participant ($\chi^2(2) = 1.20, p = .549$).

\newpage\subsection{Self-reported Attitude (T2)}
\label{app:dqi-attitude-t2-table}
\begin{table}[h!]
\centering
\scriptsize
\caption{Treatment effects on attitude and survey items (T2: After Reflection, Before Discussion). Point estimates ($\delta$) are covariate-adjusted mean differences (scale points) with 95\% confidence intervals, standardized effect sizes (Cohen's $d$), and unadjusted two-sided $p$-values ($p$).}
\label{tab:attitude_dimensions_t2_all}
\begin{tabular}{l l ccc}
\toprule
 & & \multicolumn{3}{c}{\textbf{All Participants}} \\
\cmidrule(lr){3-5}
\textbf{Dimension} & \textbf{Contrast} & $\delta$ [95\% CI] & $d$ & $p$ \\
\midrule
\textbf{Stance extremity} & AIR vs. Control & $+0.06$ $[-0.07, +0.19]$ & $+0.06$ & $.341$ \\
   & AIR vs. Static & $-0.07$ $[-0.20, +0.06]$ & $-0.07$ & $.291$ \\
   & Static vs. Control & \textbf{$+0.13^*$ $[+0.01, +0.26]$} & $+0.14$ & \textbf{$.041$} \\
\midrule
\textbf{Stance certainty} & AIR vs. Control & $+0.12$ $[-0.10, +0.34]$ & $+0.07$ & $.301$ \\
   & AIR vs. Static & $+0.01$ $[-0.20, +0.23]$ & $+0.01$ & $.912$ \\
   & Static vs. Control & $+0.10$ $[-0.11, +0.32]$ & $+0.07$ & $.334$ \\
\midrule
\textbf{Policy understanding} & AIR vs. Control & $-0.03$ $[-0.24, +0.17]$ & $-0.02$ & $.753$ \\
   & AIR vs. Static & \textbf{$+0.27^{**}$ $[+0.07, +0.47]$} & $+0.19$ & \textbf{$.010$} \\
   & Static vs. Control & \textbf{$-0.30^{**}$ $[-0.51, -0.09]$} & $-0.20$ & \textbf{$.005$} \\
\midrule
\textbf{Preparedness} & AIR vs. Control & $+0.10$ $[-0.10, +0.30]$ & $+0.07$ & $.312$ \\
   & AIR vs. Static & $+0.17$ $[-0.04, +0.37]$ & $+0.12$ & $.106$ \\
   & Static vs. Control & $-0.06$ $[-0.26, +0.14]$ & $-0.04$ & $.549$ \\
\midrule
\textbf{Outgroup Empathy} & AIR vs. Control & $+0.12$ $[-0.09, +0.33]$ & $+0.07$ & $.269$ \\
   & AIR vs. Static & \textbf{$+0.23^*$ $[+0.02, +0.44]$} & $+0.15$ & \textbf{$.033$} \\
   & Static vs. Control & $-0.11$ $[-0.32, +0.10]$ & $-0.07$ & $.305$ \\
\midrule
\textbf{Outgroup understanding} & AIR vs. Control & \textbf{$+0.24^*$ $[+0.04, +0.44]$} & $+0.16$ & \textbf{$.017$} \\
   & AIR vs. Static & \textbf{$+0.25^*$ $[+0.05, +0.45]$} & $+0.17$ & \textbf{$.017$} \\
   & Static vs. Control & $-0.01$ $[-0.21, +0.20]$ & $-0.00$ & $.947$ \\
\bottomrule
\end{tabular}
\vspace{3pt}
\par\footnotesize{\textit{Note:} $^{*}p < .05$, $^{**}p < .01$, $^{***}p < .001$ (unadjusted two-sided $p$-values).}
\end{table}

\newpage\subsection{Self-reported Attitude (T3)}
\label{app:dqi-attitude-t3-table}

\begin{table}[h!]
\centering
\scriptsize
\caption{Treatment effects on attitude and survey items (T3: After Discussion). Point estimates ($\delta$) are covariate-adjusted mean differences (scale points) with 95\% confidence intervals, standardized effect sizes (Cohen's $d$), and unadjusted two-sided $p$-values ($p$).}
\label{tab:attitude_dimensions_t3_all}
\begin{tabular}{l l ccc}
\toprule
 & & \multicolumn{3}{c}{\textbf{All Participants}} \\
\cmidrule(lr){3-5}
\textbf{Dimension} & \textbf{Contrast} & $\delta$ [95\% CI] & $d$ & $p$ \\
\midrule
\textbf{Stance extremity} & AIR vs. Control & $+0.01$ $[-0.12, +0.14]$ & $+0.01$ & $.888$ \\
   & AIR vs. Static & $-0.01$ $[-0.13, +0.12]$ & $-0.01$ & $.939$ \\
   & Static vs. Control & $+0.01$ $[-0.11, +0.14]$ & $+0.01$ & $.828$ \\
\midrule
\textbf{Stance certainty} & AIR vs. Control & $+0.04$ $[-0.15, +0.23]$ & $+0.03$ & $.684$ \\
   & AIR vs. Static & $+0.09$ $[-0.10, +0.28]$ & $+0.07$ & $.342$ \\
   & Static vs. Control & $-0.05$ $[-0.24, +0.14]$ & $-0.04$ & $.584$ \\
\midrule
\textbf{Policy understanding} & AIR vs. Control & $-0.13$ $[-0.34, +0.09]$ & $-0.09$ & $.241$ \\
   & AIR vs. Static & $+0.18$ $[-0.04, +0.39]$ & $+0.12$ & $.105$ \\
   & Static vs. Control & \textbf{$-0.31^{**}$ $[-0.52, -0.09]$} & $-0.21$ & \textbf{$.005$} \\
\midrule
\textbf{Preparedness} & AIR vs. Control & $+0.14$ $[-0.04, +0.32]$ & $+0.11$ & $.121$ \\
   & AIR vs. Static & \textbf{$+0.19^{*}$ $[+0.01, +0.37]$} & $+0.15$ & \textbf{$.039$} \\
   & Static vs. Control & $-0.05$ $[-0.23, +0.13]$ & $-0.04$ & $.604$ \\
\midrule
\textbf{Outgroup Empathy} & AIR vs. Control & $+0.24$ $[-0.01, +0.49]$ & $+0.14$ & $.059$ \\
   & AIR vs. Static & \textbf{$+0.27^{*}$ $[+0.02, +0.52]$} & $+0.15$ & \textbf{$.036$} \\
   & Static vs. Control & $-0.03$ $[-0.28, +0.22]$ & $-0.02$ & $.827$ \\
\midrule
\textbf{Outgroup understanding} & AIR vs. Control & $+0.21$ $[-0.03, +0.45]$ & $+0.13$ & $.084$ \\
   & AIR vs. Static & $+0.21$ $[-0.03, +0.45]$ & $+0.13$ & $.086$ \\
   & Static vs. Control & $+0.00$ $[-0.24, +0.24]$ & $+0.00$ & $.993$ \\
\bottomrule
\end{tabular}
\vspace{3pt}
\par\footnotesize{\textit{Note:} $^{*}p < .05$, $^{**}p < .01$, $^{***}p < .001$ (unadjusted two-sided $p$-values).}
\end{table}

\newpage\subsection{Other Survey Responses}
\begin{figure*}[h!]
  \centering
  \includegraphics[width=\textwidth]{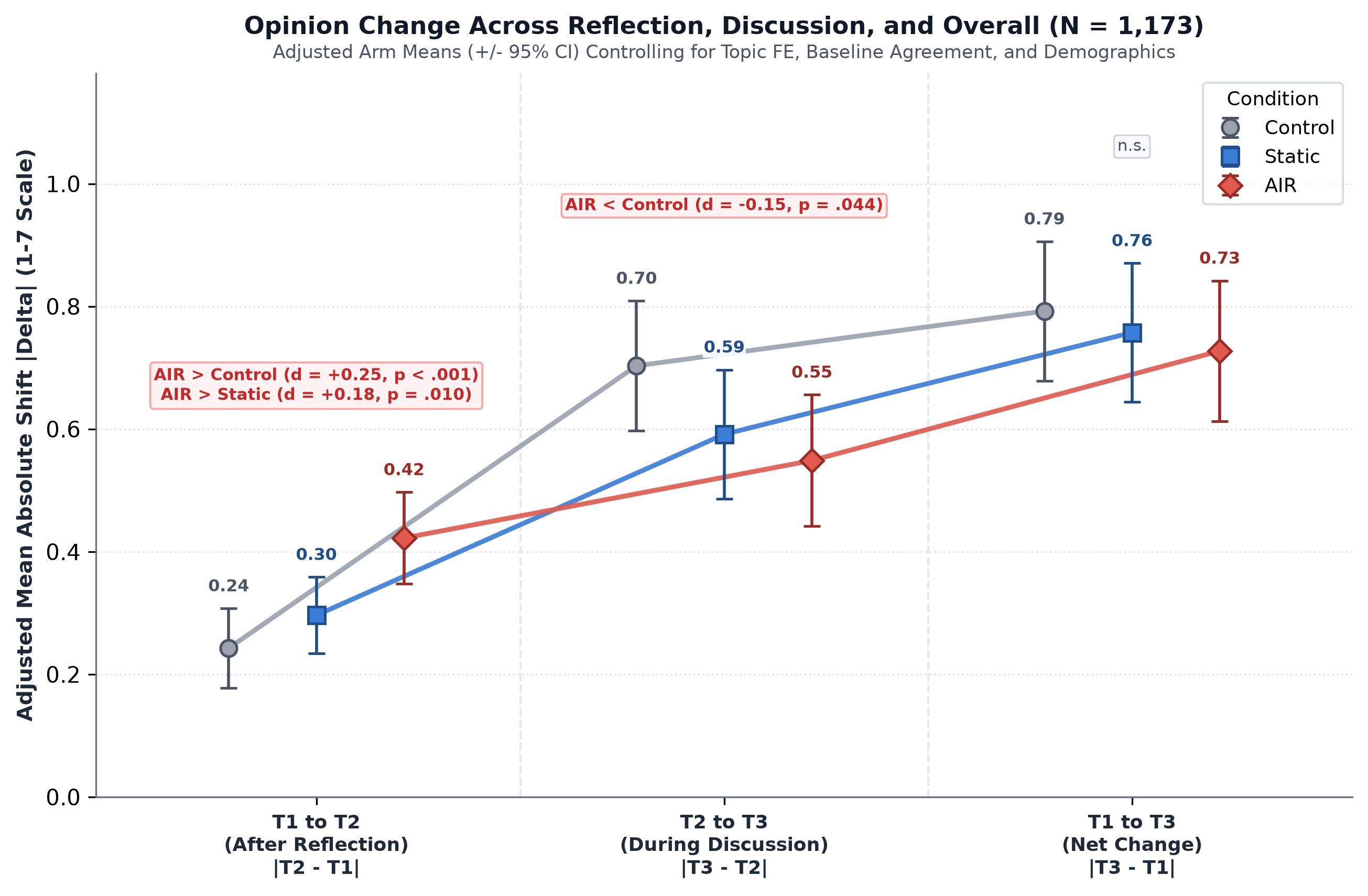}
\caption{\textbf{Magnitude of opinion change.}
Points and error bars represent covariate-adjusted condition means ($\pm 95\%$ CIs) for the absolute shift in policy stance ($|\Delta|$, 1--7 scale) after individual reflection ($|T_2 - T_1|$), during group discussion ($|T_3 - T_2|$), and net overall ($|T_3 - T_1|$).}
\label{fig:opinion_change_3intervals}
\end{figure*}

\begin{figure*}[h!]
  \centering
  \includegraphics[width=0.6\textwidth]{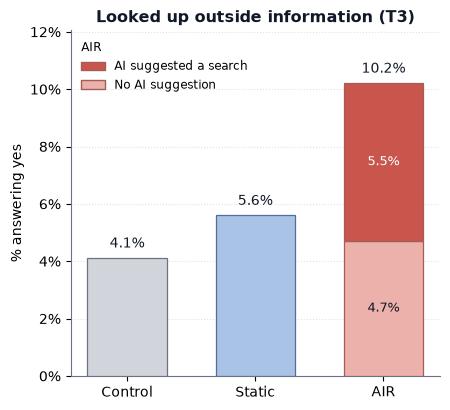}
\caption{\textbf{Looked up outside information.}
  At the end of the study, participants were asked if they had sought outside information. A higher proportion of AIR participants responded "yes". This behavior was guided by the AI Reflector’s design: AIR facilitates active reflection and prompts participants to independently seek out information ("AI suggested a search") when they directly ask for information. }
  \label{fig:outside_search}
\end{figure*}

\begin{figure*}[h!]
  \centering
  \includegraphics[width=\textwidth]{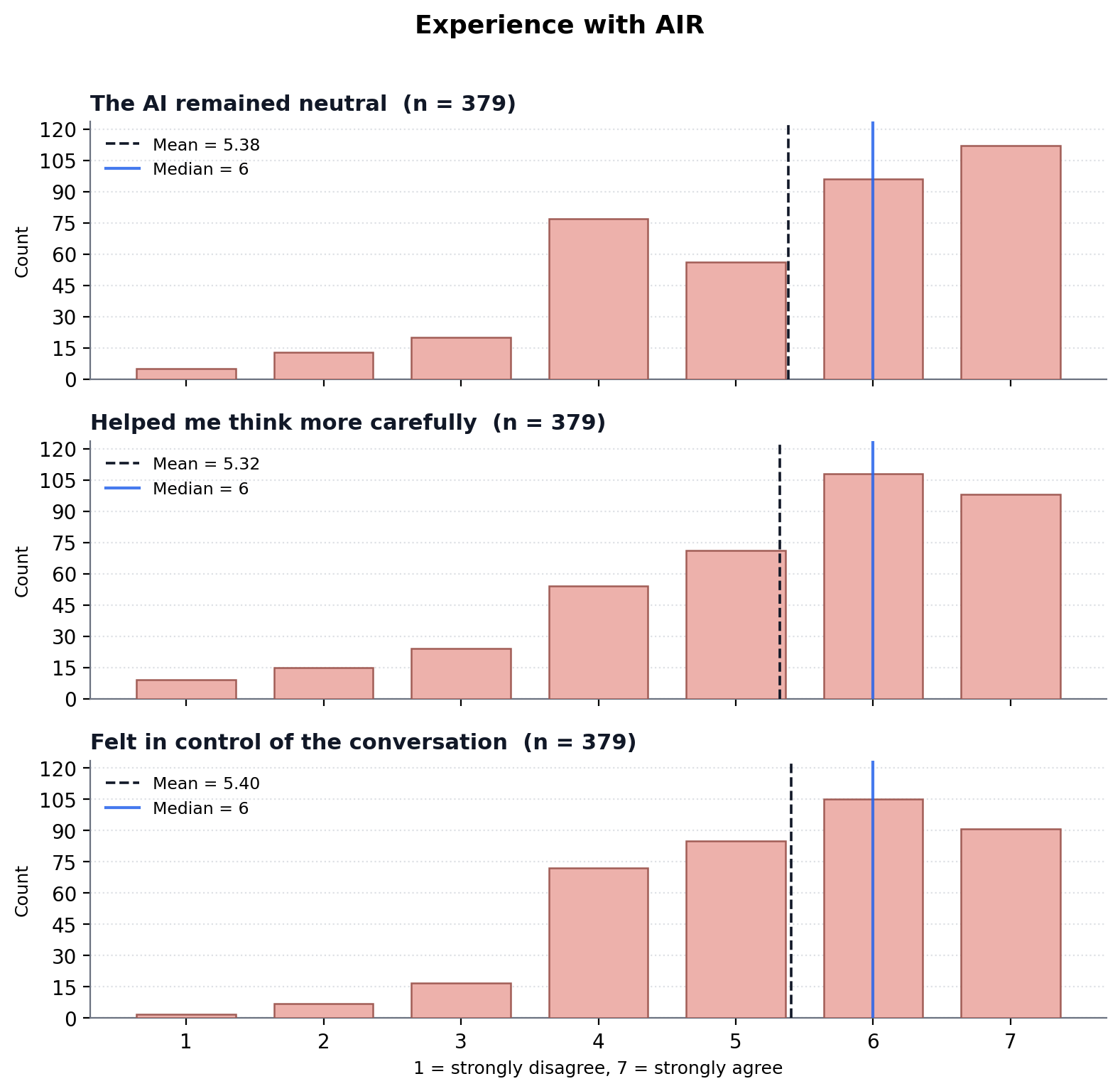}
\caption{\textbf{Experience with AIR.}
   At the end of the study participants from the AIR condition rated their reflection conversation with AIR on 7 point Likert scale (1 = strongly disagree, 7 = strongly agree). Dashed lines mark the mean; solid lines mark the median. Most participants think AIR is neutral, think it helped them think more carefully, and feel in control of the conversation.}
  \label{fig:experience-air}
\end{figure*}
\newpage\section{Qualitative Examples of Reflection and Group Deliberation}
\label{app:qualitative_examples}

To illustrate how interactive AIR versus Static reflection shaped subsequent group deliberation, Tables~\ref{tab:qualitative_pair1} and~\ref{tab:qualitative_pair2} present two matched pairs of participant trajectories drawn from the full study transcripts. 

In \textbf{Pair 1 (Medicaid Work Requirements; Table~\ref{tab:qualitative_pair1})}, both participants explicitly admitted a gap in their policy knowledge during reflection. In the Static condition, the checklist prompted the participant to write down unanswered factual questions without resolving them ($T_2$ policy understanding $= 3/7$); the participant entered the group chat still asking what the law entails and relied on unelaborated assertions ($\text{Total DQI} = 4.50$). In the AIR condition, when AIR probed how the mandate would operate in practice, the participant recognized that their support rested on ``abstract arguments'' without knowing how the policy could be implemented. By prompting the participant to connect their own premise (``health is certainly the greatest progenitor of wealth'') to the mandate's requirements, AIR surfaced the ``circular'' tension between needing healthcare to work and working to retain healthcare. In the subsequent group discussion, the participant acknowledged how the mandate could unfairly harm those with health issues and distinguished the policy's abstract principle from its practical implementation problems ($\text{Total DQI} = 6.23$, $T_1 \to T_3$ agreement: $5 \to 4$).

In \textbf{Pair 2 (Live Facial Recognition Technology; Table~\ref{tab:qualitative_pair2})}, both participants started with the same strong opposition ($T_1$ agreement $= 1$), grounded in the same core concern over public privacy and the inability to consent or opt out of facial scanning. The Static participant rehearsed these points in a solitary paragraph and subsequently contributed short, one-sided anecdotes in the group chat ($\text{Total DQI} = 3.80$). By contrast, AIR prompted the participant to weigh their privacy and consent concerns against the opposing public-safety benefits of identifying suspects and aiding law enforcement. In the resulting group discussion, despite maintaining their strong opposition ($1 \to 1 \to 1$), the AIR participant explicitly balanced their concerns over lack of consent and Flock camera abuse against the technology's utility for capturing criminals and locating missing people, while reflecting on what protections would be required ($\text{Total DQI} = 7.40$).

\begin{table*}[htbp]
\centering
\caption{\textbf{Pair 1 (Medicaid Work Requirements): Handling Knowledge Gaps and Scaffolding ``Deliberation Within.''} Trajectories through pre-discussion reflection ($T_1 \to T_2$), to synchronous group deliberation ($T_2 \to T_3$). Quotes are verbatim transcripts (minor typos preserved; ellipses indicate abridged turns).}
\label{tab:qualitative_pair1}
\footnotesize
\renewcommand{\arraystretch}{1.3}
\begin{tabularx}{\textwidth}{
  >{\raggedright\arraybackslash}p{0.2\textwidth}
  >{\raggedright\arraybackslash}X
  >{\raggedright\arraybackslash}p{0.29\textwidth}
}
\toprule
\textbf{Condition \& Trajectory} & \textbf{Reflection Process ($T_1 \to T_2$)} & \textbf{Resulting Group Discussion ($T_2 \to T_3$)} \\
\midrule
\textbf{Static Reflection}\newline
\vspace{2pt}
$T_1 \to T_2 \to T_3$\newline
Agree: $5 \to 4 \to 3$\newline
Understand ($T_2$): $3/7$\newline
Opp.\ Reasons ($T_2$): $4/7$\newline
\vspace{2pt}
\textbf{Total DQI: 4.50}
&
\textbf{Solitary Scratchpad Continuation:} ``I do not [know] enough about the policy to make a truly informed opinion. I worry about the abuse of the program from not only the recipeants [sic] but also by the states and federal govenments running the program. Im sure people will say that no [o]ne gets free handouts so they need to work but is this always the ca[s]e? will people slip through the cracks? Plenty of states and politicians have abused medicaid and took [m]oney away from the ones that need it. Is that what this is? How is this policy different than before?''\newline
\vspace{2pt}
\textit{[Unanswered factual questions remain unresolved prior to chat.]}
&
\textbf{Turn 1:} ``My question is what was the policy previously? How much of a change is this?''\newline
\textbf{Turn 10:} ``exactly. there are so many factors to consider. I suppose I should actually read the law. I only hear the talking heads yelling on both sides in the headlines.''\newline
\textbf{Turn 16:} ``keeping your health tied to your job is so insedious [sic]. Keep working and take your low pay or no healthcare for you...'' \\
\midrule
\textbf{AIR Reflection}\newline
\vspace{2pt}
$T_1 \to T_2 \to T_3$\newline
Agree: $5 \to 5 \to 4$\newline
Understand ($T_2$): $3/7$\newline
Opp.\ Reasons ($T_2$): $4/7$\newline
\vspace{2pt}
\textbf{Total DQI: 6.23}
&
\textbf{User (Turn 3):} ``my reasons are not set in stone, but based on general principles that people who want assistance from the government should be happy to contribute to society, whether this this is a beneficial opinion would require information and data''\newline
\textbf{AIR (Turn 4):} ``Since you're thinking about this in terms of contributing to society, how do you envision this requirement working in practice? What would the process look like for someone to maintain their eligibility?''\newline
\textbf{User (Turn 5):} ``In all honesty, I really couldn't tell you the most efficient way of implementing this policy, simply that my position was the most sound in an abstract space, making abstract arguments''\newline
\textbf{User (Turn 11):} ``health is certainly the greatest progenitor of wealth''\newline
\textbf{AIR (Turn 12):} ``Considering that health can be a prerequisite for work, how does that interact with your principle that people should contribute to society in order to receive government assistance? ...''\newline
\textbf{User (Turn 13):} ``I had to think about this, the feeling I get is that the question seems to be somewhat circular and in abstraction, how cna [sic] you work without health, but if you have health the logic seems to be that you must be willing to work''
&
\textbf{Turn 13:} ``\textbf{so some people who had health issues who couldn't work the policy may unfairly hurt them} ... generally the idea is that people are contributing to society if possible''\newline
\textbf{Turn 20:} ``\textbf{We can also agree that the logic seems to be sound but the implementation of the logic may have several problems}''\newline
\textbf{Turn 35:} ``\textbf{I suspect the difficulty in implementation may actually be the biggest reason this is such a logjam}'' \\
\bottomrule
\end{tabularx}
\end{table*}

\begin{table*}[htbp]
\centering
\caption{\textbf{Pair 2 (Live Facial Recognition Technology): Steering Identical Initial Concerns into Balanced Justifications.} Trajectories through pre-discussion reflection ($T_1 \to T_2$), to synchronous group deliberation ($T_2 \to T_3$). Quotes are verbatim transcripts (minor typos preserved; ellipses indicate abridged turns).}
\label{tab:qualitative_pair2}
\footnotesize
\renewcommand{\arraystretch}{1.3}
\begin{tabularx}{\textwidth}{
  >{\raggedright\arraybackslash}p{0.2\textwidth}
  >{\raggedright\arraybackslash}X
  >{\raggedright\arraybackslash}p{0.29\textwidth}
}
\toprule
\textbf{Condition \& Trajectory} & \textbf{Reflection Process ($T_1 \to T_2$)} & \textbf{Resulting Group Discussion ($T_2 \to T_3$)} \\
\midrule
\textbf{Static Reflection}\newline
\vspace{2pt}
$T_1 \to T_2 \to T_3$\newline
Agree: $1 \to 2 \to 1$\newline
Understand ($T_2$): $3/7$\newline
Opp.\ Reasons ($T_2$): $5/7$\newline
\vspace{2pt}
\textbf{Total DQI: 3.80}
&
\textbf{Solitary Scratchpad Continuation:} ``My current view is I am strongly against it and there is no way to opt out of this program. My main concern is the loss of privacy and constant surveillance... We have seen numerous cases of police abusing Flock cameras to spy on their loved ones... and also police tracking down women over state lines who are traveling for abortions. My knowledge of the scope of policy is what I read on the news... I am curious what databases are used and what is done in cases of mistaken identity. As for opposite views I'm sure it helps find missing persons or criminals on the run and solve crimes.''
&
\textbf{Turn 3:} ``What if I'm misidentified as a criminal?''\newline
\textbf{Turn 6:} ``I only have experience with flock cameras and law enforcement is abusing it tracking women over state lines for miles bc they got an abortion no way''\newline
\textbf{Turn 8:} ``In my town its local police stalking their wives and exes'' \\
\midrule
\textbf{AIR Reflection}\newline
\vspace{2pt}
$T_1 \to T_2 \to T_3$\newline
Agree: $1 \to 1 \to 1$\newline
Understand ($T_2$): $5/7$\newline
Opp.\ Reasons ($T_2$): $5/7$\newline
\vspace{2pt}
\textbf{Total DQI: 7.40}
&
\textbf{User (Turn 1 \& 3):} ``I believe that local law enforcement agencies should not be allowed to use live facial recognition technology in public spaces. It is a breach of privacy and the individuals are not giving consent... This technology scans and records individuals in public spaces. People have the right to privacy and to not be recorded without their consent.''\newline
\textbf{AIR (Turn 4):} ``If someone supported this policy, they might argue that the ability to quickly identify suspects in a crowd is a necessary trade-off for public safety. How do you think someone with that view would reason through the privacy concerns you mentioned?''\newline
\textbf{User (Turn 7):} ``I would like to clarify that I do understand the useful aspects of this technology for public safety and helping law enforcement. However, privacy is a right everyone should have, and the technology can be used for the wrong reasons as well.''
&
\textbf{Turn 12:} ``I disagree with this policy because of privacy and consent concerns... \textbf{I understand that this technology can be beneficial for law enforcement; there is also potential for misuses. This has been seen with flock cameras.} I believe the privacy risks and lack of consent are concerns that shouldn't be overlooked.''\newline
\textbf{Turn 17:} ``\textbf{The technology can be useful for capturing criminals and missing people} but once an agency has access to this technology they can use it for anything.''\newline
\textbf{Turn 31:} ``I personally have not but I have heard of flock cameras being used by officers to track down women.'' \\
\bottomrule
\end{tabularx}
\end{table*}
\newpage\section{Post-hoc Analyses}
\subsection{Knowledge-Gap Admission, Perceived Policy Understanding, and Deliberation Quality}
We investigated the incidence of participants explicitly admitting to gaps in their own knowledge during reflection (in either Static or AIR conditions), and how this behavior correlates with self-reported policy understanding and subsequent deliberation quality. While both conditions share the reflection goals, the Static condition shows instructions as a checklist that asks participants to write down what they know versus what they do not know about the policy, while AIR probes this interactively.

We coded each participant's reflection text for explicit first-person admissions of lack of policy knowledge. Static reflection participants explicitly admitted a knowledge gap at nearly twice the rate of AIR participants ($41.7\%$ vs.\ $21.5\%$). A potential reason is that AIR decided when to probe a participant's knowledge gap dynamically and therefore did not do so all the time, while every Static reflection participant was prompted to write down their knowledge gap.

We found admitting a knowledge gap strongly predicted lower post-reflection reported policy understanding at T2 (regression controls for topic, experimental conditions, $T_1$ stance, reflection length, and demographics), holding within both the AI ($\beta = -0.77, p =0.001$) and Static ($\beta = -0.49, p =0.007$) conditions. Relative to the control (mean $M$=4.91), participants who \textit{did not} admit a knowledge gap reported undiminished understanding in both AI ($M = 5.08$) and Static ($M = 4.95$) reflection, whereas those who \textit{did} admit a knowledge gap drove the entire reduction in both arms ($M = 4.20, \beta = -0.51, p = 0.006$ in AI; $M = 4.26, \beta = -0.64, p < 0.001$ in Static). However, holding topic, experimental conditions, $T_1$ stance, reflection length, and demographics constant, explicitly admitting a knowledge gap did not significantly predict overall DQI ($\beta = -0.129, p = 0.12$), although it was associated with slightly lower \textit{Level of Justification} ($\beta = -0.081, p = 0.015$). 

Overall, the pattern suggests that it is mainly participants admitting a knowledge gap reporting lower policy understanding, and static reflection prompts more participants to do so. Yet it does not automatically translate into better deliberation quality when talking to others. 

\subsection{Echo Chambers in One-Sided Groups, and Absence of Backfiring in Two-Sided Groups}
\label{app:one_vs_two_sided_extremity}
Examining changes in individual stance extremity ($|\text{Agreement} - 4|$, on a $0$--$3$ folded scale) during the group discussion ($T_2 \to T_3$) provides evidence of an ``echo chamber'' effect in one-sided groups alongside no net backfiring in two-sided groups. In one-sided cohorts, group discussion pushed individual stance extremity sharply upward overall ($M_{T2} = 2.08$ to $M_{T3} = 2.39$, $\Delta_{T2 \to T3} = +0.32$, $p<0.001$)—an increase $+0.23$ scale points larger than in two-sided cohorts (adjusting for topic, baseline stance, and demographics). This one-sided extremity increase was statistically significant within every condition, but was directionally smallest in AIR ($\Delta_{T2 \to T3} = +0.24$, $p <0.001$; cumulative $\Delta_{T1 \to T3} = +0.34$, $p < .001$) compared to Static ($\Delta_{T2 \to T3} = +0.35$, $p <0.001$; cumulative $\Delta_{T1 \to T3} = +0.45$, $p < .001$) and Control ($\Delta_{T2 \to T3} = +0.36$, $p<0.001$; cumulative $\Delta_{T1 \to T3} = +0.38$, $p = <0.001$), though pairwise between-arm contrasts within one-sided groups did not reach statistical significance ($\text{AIR vs.\ Static}: \delta = -0.09$, $p = .45$; $\text{AIR vs.\ Control}: \delta = -0.12$, $p = .30$). Conversely, in two-sided cohorts, exposure to disagreeing peers produced no net backfiring or extremity polarization during group discussion ($M_{T2} = 1.92$ to $M_{T3} = 1.96$, $\Delta_{T2 \to T3} = +0.04$, $p = .17$), a null effect that held across all three arms (AIR: $\Delta_{T2 \to T3} = +0.06$, $p = .16$; Static: $\Delta_{T2 \to T3} = -0.04$, $p = .39$; Control: $\Delta_{T2 \to T3} = +0.09$, $p = .051$; $\text{AIR vs.\ Control}: \delta = -0.03$, $p = .69$; $\text{AIR vs.\ Static}: \delta = +0.11$, $p = .10$). Together, these findings indicate that cross-cutting peer deliberation prevents attitude extremization without triggering net backfire, whereas deliberating exclusively with like-minded peers consistently amplifies stance extremity across conditions.

\newpage\section{Empirical Consistency with Prior Work}
Our work builds on a body of established theoretical and empirical frameworks in cognitive and political psychology, with which we find noteworthy alignment.

First, consistent with argumentative theory of reasoning \citep{mercier2011humans, mercier2012reasoning}, we observe that the static reflection condition led to an increase in stance extremity. Importantly, this observation is limited to the Static condition, showing that self-reflection is not bound to these negative outcomes. In particular, we find that AIR may support productive self-reflection, consistently with Goodin's ``deliberation within'' suggestion \citep{goodin2003does}.

Consistently with \cite{fernbach2013political}'s work on IOED, we find that self-reflection by enumerating reasons had a small effect on IOED reduction. Our findings further align with \cite{fernbach2013political}'s exploratory results on evaluative reason-generation increasing stance extremity (although they find a null effect on depolarization overall). \footnote{Note that both work raise an unanswered question pertaining to the mechanism through which static self-reflection lead both to decreased policy understanding (which may be seen as a decrease in overconfidence) and higher stance extremity.} Additionally, the analysis of the free-form reflection texts reveals that IOED reduction in our study is primarily driven by the subset of participants who explicitly acknowledge gaps in their understanding, as in \cite{fernbach2013political}.

Last, participants in the AIR condition reported seeking out external information significantly more often than participants in the other groups (see \Cref{fig:outside_search}), since AIR is instructed to nudge participants to gather outside information themselves rather than simply providing information from its own knowledge. Among the 53 participants who were prompted to look up information themselves, 22 (41.51\%) of them reported searching. 

These AIR searchers produced the largest opinion shift of any phase (+0.76, 47.4$\%$ move)—larger than peer deliberation for either searchers (+0.53) or non-searchers (+0.55) changes in stance. This observation is consistent with Goodin's theoretical and empirical work on ``deliberation within,'' which suggested that most of opinion change was happening when deliberators gathering new information before actually deliberating. It should however be interpreted cautiously given the small sample size.

Overall, this work provides a unifying framework between these theories: static self-reflection worsens overall deliberation quality and preparedness as suggested by \citep{landemore2010talking}, but their critique of ``deliberation within'' is not applicable when reflection, as enabled by AIR, allows participants to widen their views. 
\end{document}